\documentclass[10pt,english]{article}
\usepackage[latin9]{inputenc}
\usepackage{geometry}
\usepackage{xcolor}
\usepackage{pdfcolmk}
\usepackage{array}
\usepackage{float}
\usepackage{amsmath}
\usepackage{graphicx}
\PassOptionsToPackage{normalem}{ulem}
\usepackage{ulem}

\makeatletter

\providecommand{\tabularnewline}{\\}
\newcommand{\lyxdot}{.}

\providecolor{lyxadded}{rgb}{0,0,1}
\providecolor{lyxdeleted}{rgb}{1,0,0}

\DeclareRobustCommand{\lyxsout}[1]{\ifx\\#1\else\sout{#1}\fi}

\newcommand{\lyxaddress}[1]{
	\par {\raggedright #1
	\vspace{1.4em}
	\noindent\par}
}

\usepackage{multicol}
\usepackage{cite}
\usepackage{nameref,hyperref}
\usepackage{lineno}

\newif\ifComments
\Commentstrue

\makeatother

\usepackage{babel}
\begin{document}
\title{Paradoxical phase response of gamma rhythms facilitates their entrainment
in heterogeneous networks}
\author{Xize Xu, Hermann Riecke}
\maketitle

\lyxaddress{\center{Department of Engineering Science and Applied Mathematics, \\ Northwestern
University, Evanston, IL 60208, USA}}
\begin{abstract}
The synchronization of different $\gamma$-rhythms arising in different
brain areas has been implicated in various cognitive functions. Here,
we focus on the effect of the ubiquitous neuronal heterogeneity on
the synchronization of PING (pyramidal-interneuronal network gamma)
and ING (interneuronal network gamma) rhythms. The synchronization
properties of rhythms depends on the response of their collective
phase to external input. We therefore determined the macroscopic phase-response
curve for finite-amplitude perturbations (fmPRC), using numerical
simulation of all-to-all coupled networks of integrate-and-fire (IF)
neurons exhibiting either PING or ING rhythms. We show that the intrinsic
neuronal heterogeneity can qualitatively modify the fmPRC. While the
phase-response curve for the individual IF-neurons is strictly positive
(type I), the fmPRC can be biphasic and exhibit both signs (type II).
Thus, for PING rhythms, an external excitation to the excitatory cells
can, in fact, delay the collective oscillation of the network, even
though the same excitation would lead to an advance when applied to
uncoupled neurons. This paradoxical delay arises when the external
excitation modifies the internal dynamics of the network by causing
additional spikes of inhibitory neurons, whose delaying within-network
inhibition outweighs the immediate advance caused by the external
excitation. These results explain how intrinsic heterogeneity allows
the PING rhythm to become synchronized with a periodic forcing or
another PING rhythm for a wider range in the mismatch of their frequencies.
We demonstrate a similar mechanism for the synchronization of ING
rhythms. Our results identify a potential function of neuronal heterogeneity
in the synchronization of coupled $\gamma$-rhythms, which may play
a role in neural information transfer via communication through coherence.
\end{abstract}
\textbf{Author Summary} 

The interaction of a large number of oscillating units can lead to
the emergence of a collective, macroscopic oscillation in which many
units oscillate in near-unison or near-synchrony. This has been exploited
technologically, e.g., to combine many coherently interacting, individual
lasers to form a single powerful laser. Collective oscillations are
also important in biology. For instance, the circadian rhythm of animals
is controlled by the near-synchronous dynamics of a large number of
individually oscillating cells. In animals and humans brain rhythms
reflect the coherent dynamics of a large number of neurons and are
surmised to play an important role in the communication between different
brain areas. To be functionally relevant, these rhythms have to respond
to external inputs and have to be able to synchronize with each other.
We show that the ubiquitous heterogeneity in the properties of the
individual neurons in a network can contribute to that ability. It
can allow the external inputs to modify the internal network dynamics
such that the network can follow these inputs over a wider range of
frequencies. Paradoxically, while an external perturbation may delay
individual neurons, their ensuing within-network interaction can overcompensate
this delay, leading to an overall advance of the rhythm.

\section{Introduction}

Collective oscillations or rhythms representing the coherent dynamics
of a large number of coupled oscillators play a significant role in
many systems. In the technological realm they range from laser arrays
and Josephson junctions to micromechanical oscillators \cite{BrJo05a,WiCo96}.
Among the important biological examples are the heart rhythm, the
circadian rhythm generated by the suprachiasmatic nucleus \cite{LiWe97},
the segmentation clock controlling the somite formation during development
\cite{VenOat20}, and brain waves \cite{Wa10}. One prominent brain
rhythm is the widely observed $\gamma$-rhythm with frequencies in
the range 30-100Hz. The coherent spiking of the neurons underlying
this rhythm likely enhances the downstream impact of the neurons participating
in the rhythm. The rhythmic alternation of low and high activity has
been suggested to play a significant role in the communication between
different brain areas \cite{BoEp05,BoKo08}. That communication has
also been proposed to be controled by the coherence of the rhythms
in the participating brain areas \cite{BosFri12,RobDeW13,BuzSch15,Fri15,PalBat17,DumGut19}.

For collective oscillations or rhythms to play a constructive role
in a system they need to respond adequately to external perturbations
and stimuli. For instance, for the circadian rhythm it is essential
that it can be reliably entrained by light and phase-lock to its daily
variation. Similarly, if rhythms are to play a significant role in
the communication between different brain areas, their response to
input from other areas represents a significant determinant of their
function. Moreover, the stimulation and entrainment of $\gamma$-rhythms
by periodic sensory input is being considered as a therapeutic approach
for some neurodegenerative diseases \cite{AdaTsa19}.

Even small perturbations can affect oscillations significantly in
that they can advance or delay the oscillations, i.e. they can change
the phase of the oscillators. This change typically depends not only
on the strength of the perturbation but, importantly, also on the
timing of the perturbations and is expressed in terms of the phase
response curve (PRC), which has been studied extensively for individual
oscillators \cite{SchBut12}. For infinitesimal perturbations the
PRC can be determined elegantly using the adjoint method \cite{BrMo04}.

If the collective oscillation of a network of interacting oscillators
is sufficiently coherent, that system can be thought of as a single
effective oscillator. Consequently, the response of the macroscopic
phase of the collective oscillation to external perturbations and
the mutual interaction of multiple collective oscillations is of interest.
The macroscopic phase-response curve (mPRC) has been obtained in various
configurations, including noise-less heterogeneous phase oscillators
\cite{KaNa10a,LePi10}, noisy identical phase oscillators \cite{KaNa08,KaNa10},
noisy excitable elements \cite{KawKur11}, and noisy oscillators described
by the theta-model \cite{KotErm14}, which is equivalent to the quadratic
integrate-fire model for spiking neurons. Recent work has used the
reduction of networks of quadratic integrate-fire neurons to two coupled
differential equations for the firing rate and the mean voltage \cite{MonRox15},
which is related to the Ott-Antonsen theory \cite{OtAn08,LukSo13},
to develop a method to obtain the infinitesimal macroscopic PRC (imPRC)
for excitatory-inhibitory spiking networks \cite{HanFor15,DumGut17}.

A key difference between the response of an individual oscillator
to a perturbation and that of a collective oscillation is the fact
that the degree of synchrony of the collective oscillation can change
as a result of the perturbation, reflecting a change in the relations
between the individual oscillators. Thus, the phase response of a
collective oscillation to a brief perturbation consists not only of
the immediate change in the phases of the individual oscillators caused
by the perturbation, but includes also a change in the collective
phase that can result from the subsequent convergence back to the
phase relationship between the oscillators corresponding to the synchronized
state, which is likely to have been changed by the perturbation \cite{LePi10}.
Interestingly, it has been observed that the infinitesimal macroscopic
phase response can be qualitatively different from the phase response
of the individual elements. Thus, even if the individual oscillators
have a type-I PRC, i.e. a PRC that is strictly positive or negative,
the mPRC of the collective oscillation can be of type II, i.e. it
can exhibit a sign change as a function of the phase \cite{KawKur11,KotErm14,AkaKot18}.

Here we investigate the interplay between external perturbations and
the internal interactions among neurons in inhibitory and in excitatory-inhibitory
networks exhibiting $\gamma$-rhythms of the ING- and of the PING-type.
We focus on networks comprised of neurons that are not identical,
leading to a spread in their individual phases and a reduction in
the degree of their synchrony. How does this phase dispersion affect
the response of the macroscopic phase of the rhythm to perturbations?
Does it modify the ability of the network to follow a periodic perturbation
?

We show that the dispersion in the phase together with the within-network
interactions among the neurons can be the cause of a paradoxical phase
response: an external perturbation that\emph{ delays} each individual
neuron can \emph{advance} the macroscopic rhythm. We identify the
following mechanism underlying this paradoxical response: external
perturbations that delay individual neurons sufficiently allow the
within-network inhibition generated by early-spiking neurons to suppress
the spiking of less excited neurons. This results in a reduced within-network
inhibition, which reduces the time to the next spike volley, speeding
up the rhythm. This paradoxical phase response increases with the
neuronal heterogeneity and allows the network to phase-lock to periodic
external perturbations over a wider range of detuning. Thus, the desynchronization
within the network enhances its synchronizabilty with other networks.
The mechanism is closely related to that underlying the enhancement
of synchronization of collective oscillations by uncorrelated noise
\cite{MenRie18} and the enhanced entrainment of the rhythm of a homogeneous
network to periodic input if that input exhibits phase dispersion
across the network \cite{WhBa00,SerKop13}. We demonstrate and analyze
these behaviors for networks of inhibitory neurons (ING-rhythm) and
for networks comprised of excitatory and inhibitory neurons (PING-rhythm).

\section{Results}

We investigated the impact of neuronal heterogeneity on the response
of the phase of $\gamma$-rhythms to brief external perturbations
and the resulting ability of rhythms to synchronize to periodic input.
As described in the Methods, we used networks comprised of minimal
integrate-fire neurons that interact with each other through synaptic
pulses modeled via delayed double-exponentials. To study ING-rhythms
all neurons were inhibitory, while for the PING-rhythms we used excitatory-inhibitory
networks. In both cases, the coupling within each population was all-to-all.
Throughout, we implemented the neuronal heterogeneity by injecting
a different steady bias current $I_{bias}$ into each neuron. Our
analysis suggests that the origin of the neuronal heterogeneity plays
only a minor role as long as it leads to a dispersion of their spike
times \cite{MenRie18}.

\subsubsection*{Paradoxical Phase Response of Heterogeneous Networks: ING-Rhythm}

\begin{figure}
\begin{centering}
\includegraphics[width=0.8\linewidth]{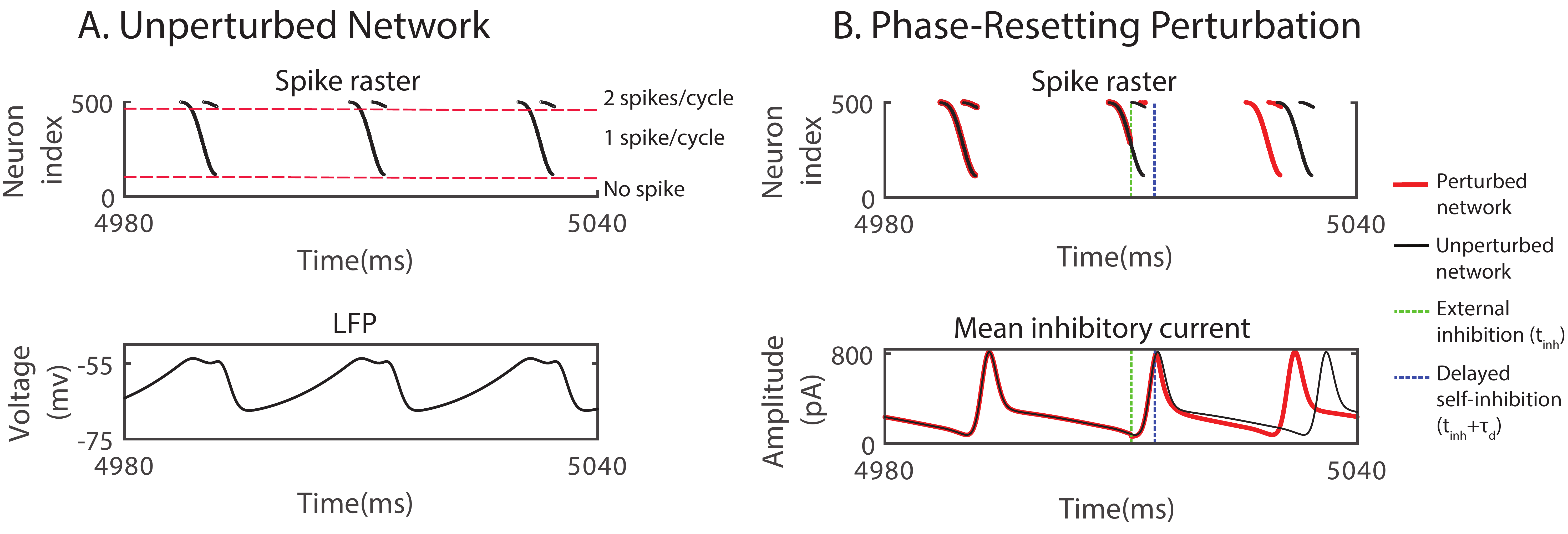}
\par\end{centering}
\caption{\textbf{ING-rhythm can be advanced by inhibition while individual
neurons are delayed.} (A) Top: spike raster of neurons spiking sequentially
in the order of their input strength (increasing with neuron index).
Bottom: mean voltage across the network (LFP). (B) External inhibition
advanced the rhythm. Top: raster plot of spikes without (black) and
with (red) external inhibitory pulse. Bottom: Average of the total
inhibitory current each neuron received from the other neurons within
the network. \label{fig:Sketch-of-model}\textbf{ }$I^{(I)}=$ 20.4
pA, $C_{V}^{(I)}$= 0.15, $f_{network}=47$ Hz. In (B), perturbations
were made with a square-wave inhibitory current pulse with duration
0.1 ms and amplitude 3200 pA to each neuron, resulting in a 4 mV rapid
hyperpolarization.}
\end{figure}

In the absence of external perturbations the all-to-all inhibition
among the neurons lead to rhythmic firing of the neurons. Due to their
heterogeneity they did not spike synchronously but sequentially, as
shown in Fig.\ref{fig:Sketch-of-model}A, where the neurons are ordered
by the strength of their bias current. The dependence of the phase
dispersion on the coefficient of variation of the heterogeneity in
the bias current (CV) is shown in Suppl. Figure \ref{fig:phase_dispersion_CV}.
For sufficiently large heterogeneity some neurons never spiked: while
the weak bias current they received would have been sufficient to
induce a spike eventually, the strong inhibition that was generated
by the neurons spiking earlier in the cycle suppressed those late
spikes. Neurons with strong bias current could spike multiple times.

A brief, inhibitory external input delivered to all neurons (green
dashed line in Fig.\ref{fig:Sketch-of-model}B) delayed each neuron.
The degree of this individual delay depended on the timing of the
input, as is reflected in the PRC of the individual neurons. If the
perturbation was applied during the time between the spike volleys,
the delay of each neuron had no further consequence and the overall
rhythm was delayed. However, if the same inhibitory perturbation arrived
during a spike volley (dashed green line in Fig.\ref{fig:Sketch-of-model}B),
it could advance the overall rhythm. As illustrated in Fig.\ref{fig:Sketch-of-model}B,
only the spiking of the late neurons was delayed by the perturbation.
Importantly, with this delay some neurons did not spike before the
within-network inhibition triggered by the early-spiking neurons (dashed
blue line in Fig.\ref{fig:Sketch-of-model}B) became strong enough
to suppress the spiking of the late neurons altogether. With fewer
neurons spiking, the all-to-all inhibition within the network was
reduced, allowing all neurons to recover earlier, which lead to a
shorter time to the next spike volley. If the speed-up was larger
than the immediate delay induced by the external inhibition, the overall
phase of the rhythm was advanced by the delaying inhibition.

\begin{figure}
\begin{centering}
\includegraphics[width=1\linewidth]{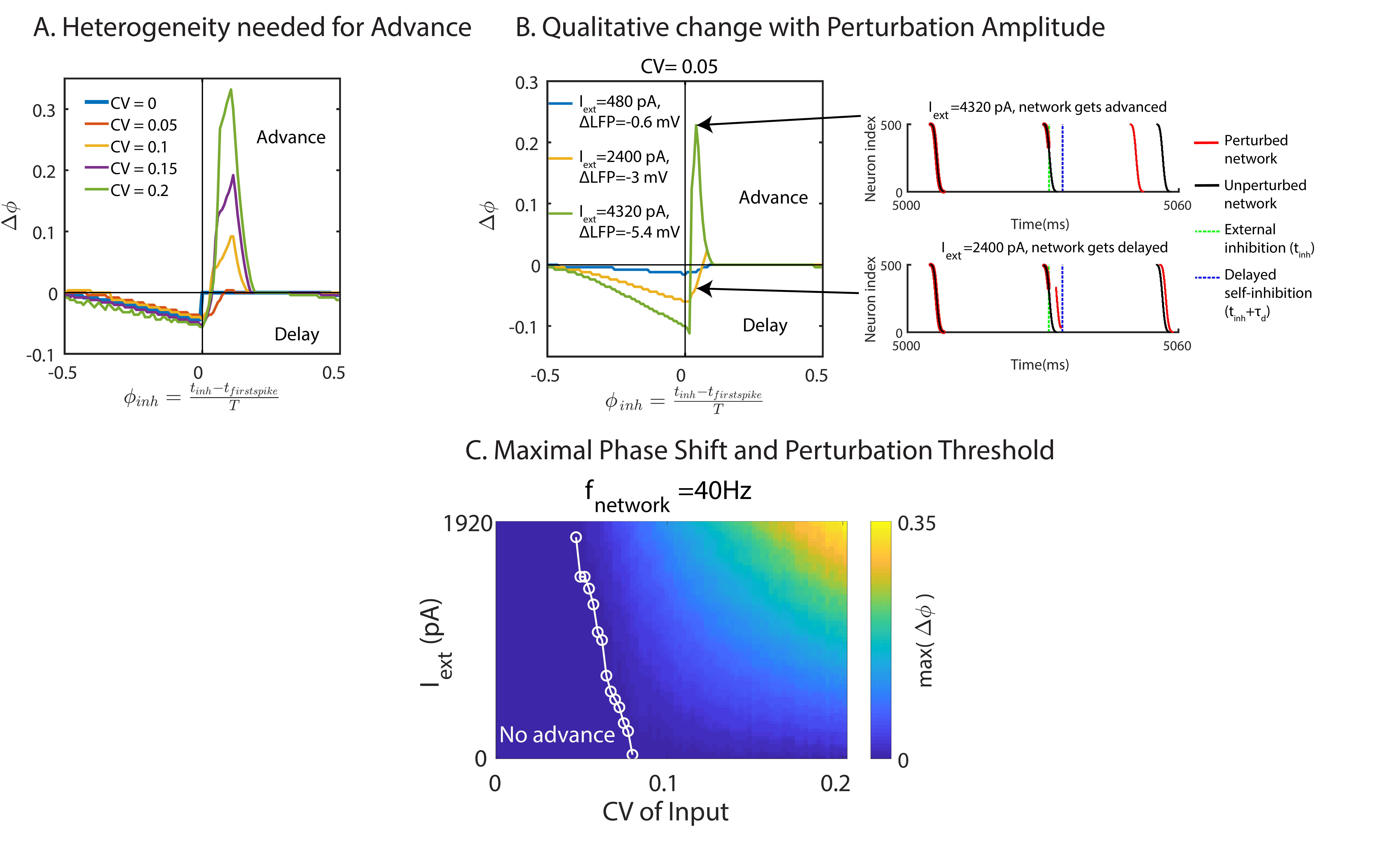}
\par\end{centering}
\caption{fmPRC of heterogeneous ING network. (A) Phase shift in response to
inhibition for different neuronal heterogeneity but fixed natural
frequency ($f_{network}=40$Hz). The paradoxical phase advance increased
with neuronal heterogeneity. (B) fmPRC changed qualitatively with
the amplitude of the perturbation. Left: fmPRC for three different
perturbation amplitudes. Right: raster plot of spikes without (black)
and with (red) external inhibition. Top: strong perturbation advanced
the network. Bottom: weak perturbation applied at the same time as
in the top figure. The network was delayed. (C) Maximal phase advance
as a function of neuronal heterogeneity and external inhibition strength.
The threshold of the inhibition amplitude to obtain an advance decreased
with heterogeneity (white line). $f_{network}$ was kept constant
($f_{network}=40$Hz). \label{fig:Heterogeneous-network-biphasic}
In (A)-(C), perturbations were made with a square-wave inhibitory
current pulse with duration 0.1 ms to each interneuron. In (A), the
amplitude of the current was 1600 pA, resulting in a 2 mV rapid hyperpolarization.}
\end{figure}

As the example in Fig.\ref{fig:Sketch-of-model}B shows, the paradoxical
phase response requires proper timing of the perturbation. We therefore
determined quantitatively the macroscopic phase-response curve (PRC)
of the rhythm. To do so we measured computationally the amount a brief
current injection shifted the phase of the rhythm (Fig.\ref{fig:Heterogeneous-network-biphasic}A).
We defined the phase as the normalized time since the first spike
in the most recent volley of spikes. Reflecting the strictly positive
PRC of the individual integrate-fire neurons, without heterogeneity
($CV=0$) external inhibition always delayed the rhythm, independent
of the timing of the pulse. In contrast, in heterogeneous networks
the rhythm could be advanced if the same inhibitory perturbation was
applied shortly after the first spikes in the spike volley\textbf{
($\phi_{inh}>0$)}.\textbf{ }Increasing the neuronal heterogeneity
enhanced this phase advance, since it shifted the within-network inhibition
driven by the leading neurons to earlier times, while it delayed the
lagging neurons. As a result, for the same external perturbation,
a larger fraction of neurons that would spike in the absence of the
external inhibition was sufficiently delayed to have their spikes
be suppressed by the within-network inhibition (cf. Fig.\ref{fig:Sketch-of-model}B),
reducing the within-network inhibition and with it the time to the
next spike volley. To keep the frequency of the unperturbed network
fixed in\textbf{ }Fig.\ref{fig:Heterogeneous-network-biphasic}A,
we reduced the tonic input with increasing heterogeneity, which enhanced
the phase advance. However, even if the tonic input was kept fixed,
the phase advance increased with heterogeneity (Suppl. Fig.\ref{fig:fixed_tonic_current}).

For weak heterogeneity the paradoxical phase response occurred only
for sufficiently strong perturbations, i.e. it did not arise in the
infinitesimal macroscopic PRC (imPRC). Thus, the phase response changed
qualitatively as the amplitude of the perturbation was strong enough
to delay the spikes of sufficiently many slow neurons until the self-inhibition
of the network set in and suppressed their spikes (Fig.\ref{fig:Heterogeneous-network-biphasic}B).
As the CV of the neurons was increased, the dispersion was large enough
that the spikes of the lagging neurons were suppressed by the self-inhibition
of the network even in the absence of an external perturbation. Above
that threshold value of CV the paradoxical phase response occurred
even for infinitesimal perturbations (Fig.\ref{fig:Heterogeneous-network-biphasic}C).

\begin{figure}
\begin{centering}
\includegraphics[width=1\linewidth]{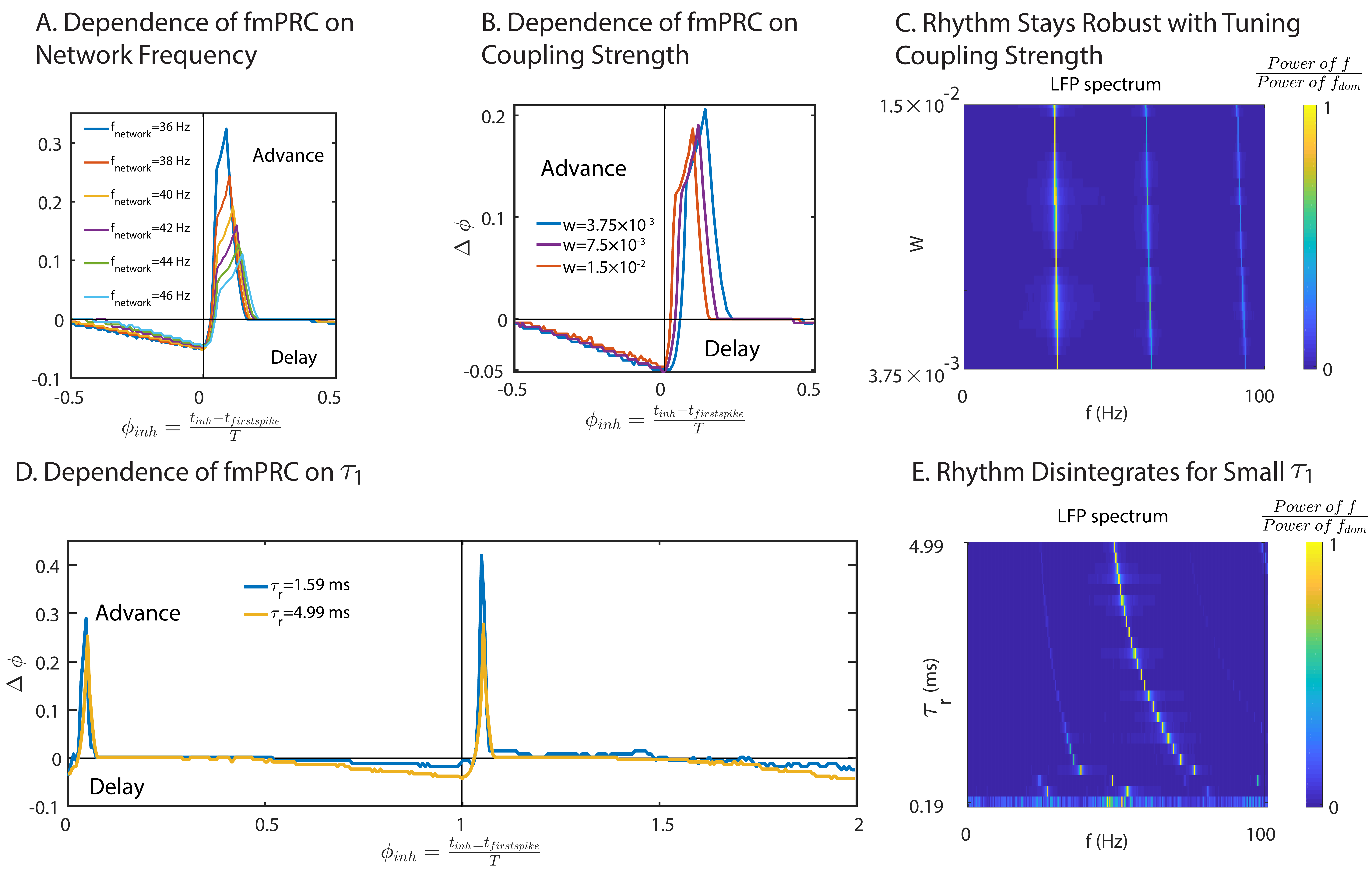}
\par\end{centering}
\caption{The paradoxical phase response of a heterogeneous ING network is robust.
(A) The phase advance of the fmPRC decreased with the natural frequency
($CV^{(I)}=0.15$). (B) The fmPRC did not depend sensitively on the
within-network coupling strength $W$ ($CV^{(I)}=0.15$, $I^{(I)}=$15.8
pA). (C) The Fourier spectrum of the LFP as a function of the coupling
strength $W$. Parameters as in (B). (D) Paradoxical phase response
in the absence of an explicit delay, $\tau_{d}=0$, for different
effective synaptic delays due to different synaptic rise times $\tau_{1}^{I}$
($CV^{(I)}=0.05$, $I^{(I)}=$15.8 pA). For low $\tau_{1}^{I}$ (blue
curve), the shift alternated in subsequent cycles reflecting the subhamornic
nature of the rhythm. (E) The Fourier spectrum of the LFP as a function
of the effective synaptic delay (synaptic time constant of rise $\tau_{1}^{I}$).
With decreasing $\tau_{1}^{I}$, a subharmonic peak emerged and eventually
the rhythm disintegrated. Parameters as in D. In (A), (B) and (D),
perturbations were made with a square-wave inhibitory current pulse
with duration 0.1 ms to each interneuron. In (A) and (B), the amplitude
of the current was 1600 pA, resulting in a 2 mV rapid hyperpolarization.
In (D), the amplitude of the current was 400 pA, resulting in a 0.5
mV rapid hyperpolarization. \label{fig:Heterogeneous-network-PRC-robustness}}
\end{figure}

The paradoxical phase response was robust with respect to changes
in the natural frequency of the network, the coupling strength, and
the effective synaptic delay, as long as the rhythm persisted. The
paradoxical phase advance increased with decreasing natural frequency
of the network, since the inhibition had a stronger effect for lower
mean input strength (Fig.\ref{fig:Heterogeneous-network-PRC-robustness}A).
Changing the within-network coupling strength by a factor of 2 up
or down did not substantially affect the paradoxical phase response
(Fig.\ref{fig:Heterogeneous-network-PRC-robustness}B) nor the strength
of the rhythm (Fig.\ref{fig:Heterogeneous-network-PRC-robustness}C).
Even without explicit synaptic delay ($\tau_{d}=0$), the effective
delay given by the double-exponential synaptic interaction was sufficient
to render a paradoxical response (Fig.\ref{fig:Heterogeneous-network-PRC-robustness}D).
However, when this effective delay was reduced by decreasing the rise
time $\tau_{1}^{I}$ of the synaptic current, the rhythm itself developed
a strong subharmonic component and eventually disintegrated (Fig.\ref{fig:Heterogeneous-network-PRC-robustness}E).
In the subharmonic regime the paradoxical phase advance alternated
in consecutive cycles of the rhythm.

In \cite{DumGut17,DumGut19} the exact reduction of all-to-all coupled
heterogeneous networks of quadratic integrate-fire neurons to 2 coupled
ordinary differential equations for each network \cite{MonRox15}
has been used to obtain the infinitesimal macroscopic phase-response
curve (imPRC) for ING and PING networks. They obtained biphasic response
only if the excitatory perturbation was applied to the population
of inhibitory neurons; for perturbations to the excitatory neurons
they found only monophasic response (type-I). This is presumably due
to the lack of a delay in the single-exponential synaptic interactions
used in \cite{DumGut17,DumGut19}.

\subsubsection*{Enhancing entrainment of ING-rhythms through network heterogeneity}

In order to allow communication by coherence \cite{Fr05,Fri15}, the
rhythms in different brain areas need to be sufficiently phase-locked
with each other. As a simplification of two interacting $\gamma$-rhythms,
we therefore investigated the ability of the rhythm in a network to
be entrained by a periodic external input, particularly focusing on
the possibly facilitating role of neuronal heterogeneity. Motivated
by the paradoxical phase response induced by the heterogeneity, we
addressed, in particular, the question whether an ING network can
be sped up by inhibition to entrain it with a faster network.

\begin{figure}
\begin{centering}
\includegraphics[width=0.7\linewidth]{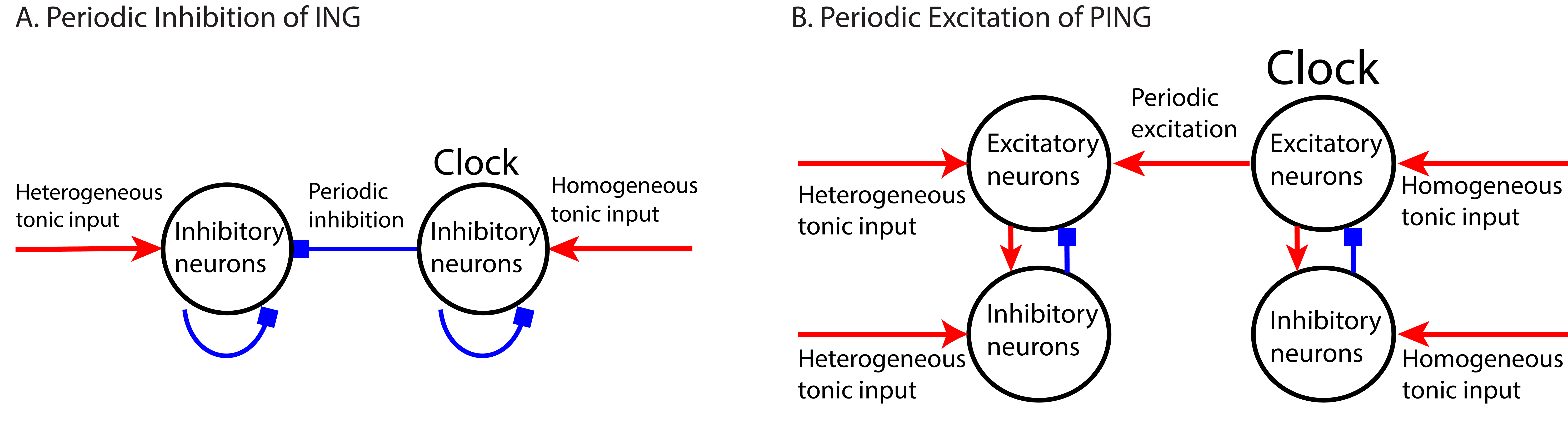}
\par\end{centering}
\caption{Sketch of computational models. (A) ING rhythm receives periodic inhibitory
input generated from another `clock' ING rhythm. (B) PING rhythm receives
periodic excitatory input by its E-population generated from another
`clock' PING rhythm. \label{fig:Sketch}}
\end{figure}

The network considered here was the same as that used to analyze the
fmPRC. The within-network interaction was an all-to-all inhibition
with synaptic delay $\tau_{d}$, resulting in a rhythm with natural
frequency $f_{natural}$, Each neuron received heterogeneous input
$I_{bias}$ and inhibitory periodic pulses with frequency $f_{clock}$
. The latter can be considered as the output of another ING-network
and were, in fact, generated that way (Fig.\ref{fig:Sketch}). We
refer to this external input as the `clock'. The detuning $\Delta f=f_{clock}-f_{natural}$
was a key control parameter\textbf{.}

\begin{figure}
\begin{centering}
\includegraphics[width=1\linewidth]{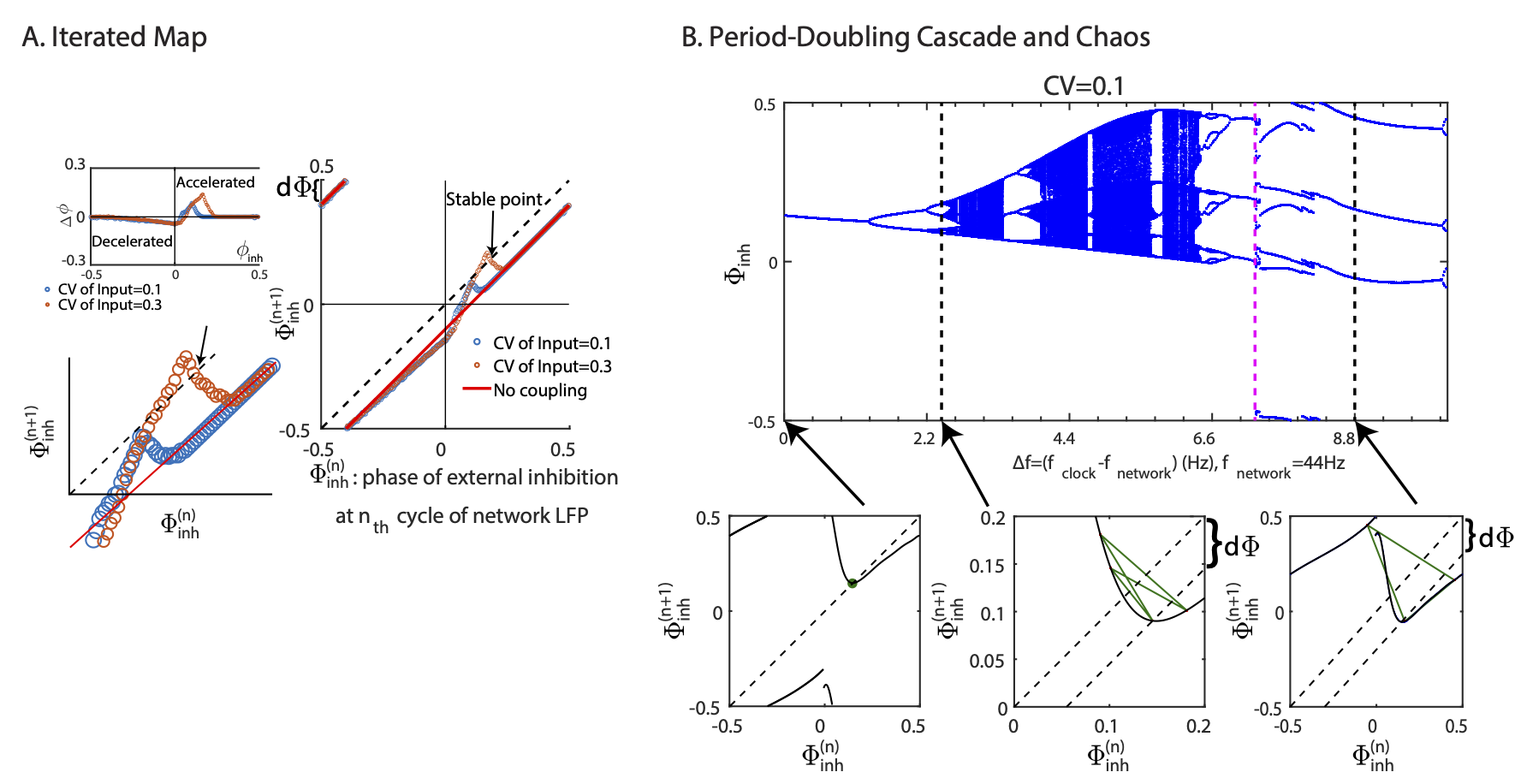}
\par\end{centering}
\caption{Connection between fmPRC and the synchronization of $\gamma$-rhythms.
(A) Iterated map of $\Phi_{inh}$. The network can be synchronized
by faster periodic inhibiton under sufficiently large advancing phase
response. (B) Top: The bifurcation diagram of the iterated map of
$\Phi_{inh}$ with varying detuning $\Delta f$. To the right of the
magenta dashed line ($\Delta f=7.28$ Hz) the attractors involve points
on both sides of the discountinuity of the map. Bottom from left to
right: iterated maps of $\Phi_{inh}$ for $\Delta f=0,2.44,8.8$ Hz.
The distance between the diagonal and subdiagonal line represents
the detuning between the network and periodic input. In (A), the fmPRC
was determined for a $\delta$-pulse perturbation, in (B) for a double-exponential
inhibitory current (cf. (\ref{eq:double_exp_1},\ref{eq:double_exp_2}))
was used as in Fig.\ref{fig:fmPRC-syn}. \label{fig:fmPRC-sketch}}
\end{figure}

For periodic input the fmPRC allows the definition of an iterated
map describing the response of the network. For periodic $\delta$-pulses
that map is shown in Fig.\ref{fig:fmPRC-sketch}A. For positive detuning,
i.e. when the clock is faster than the network, entrainment requires
that the phase response is paradoxical in order for the rhythm to
be sped up by the inhibition. If the heterogeneity and the resulting
phase response are sufficiently large, the maximum of the iterated
map crosses the diagonal, generating a stable and an unstable fixed
point. The former is the desired entrained state.

As the detuning is increased the iterated map is shifted downward.
This can decrease the slope of the iterated map at the fixed point
below -1, destabilizing the fixed point in a period-doubling bifurcation.
For periodic pulses comprised of double-exponential inhibitory currents
(cf. (\ref{eq:double_exp_1},\ref{eq:double_exp_2})) a rich bifurcation
scenario emerged (Fig.\ref{fig:fmPRC-sketch}B). Note that the full
map is not continuous and not unimodal (cf. first bottom panel of
Fig.\ref{fig:fmPRC-sketch}B). Nevertheless, for $\Delta f<7.17$
Hz the attractor remains near the unstable fixed point and displays
a period-doubling cascade to chaos and multiple periodic windows.
For $\Delta f>7.28$ Hz, however, the attractor includes points on
both sides of the discontinuity (cf. third bottom panel in Fig.\ref{fig:fmPRC-sketch}B).

\begin{figure}
\begin{centering}
\includegraphics[width=1\linewidth]{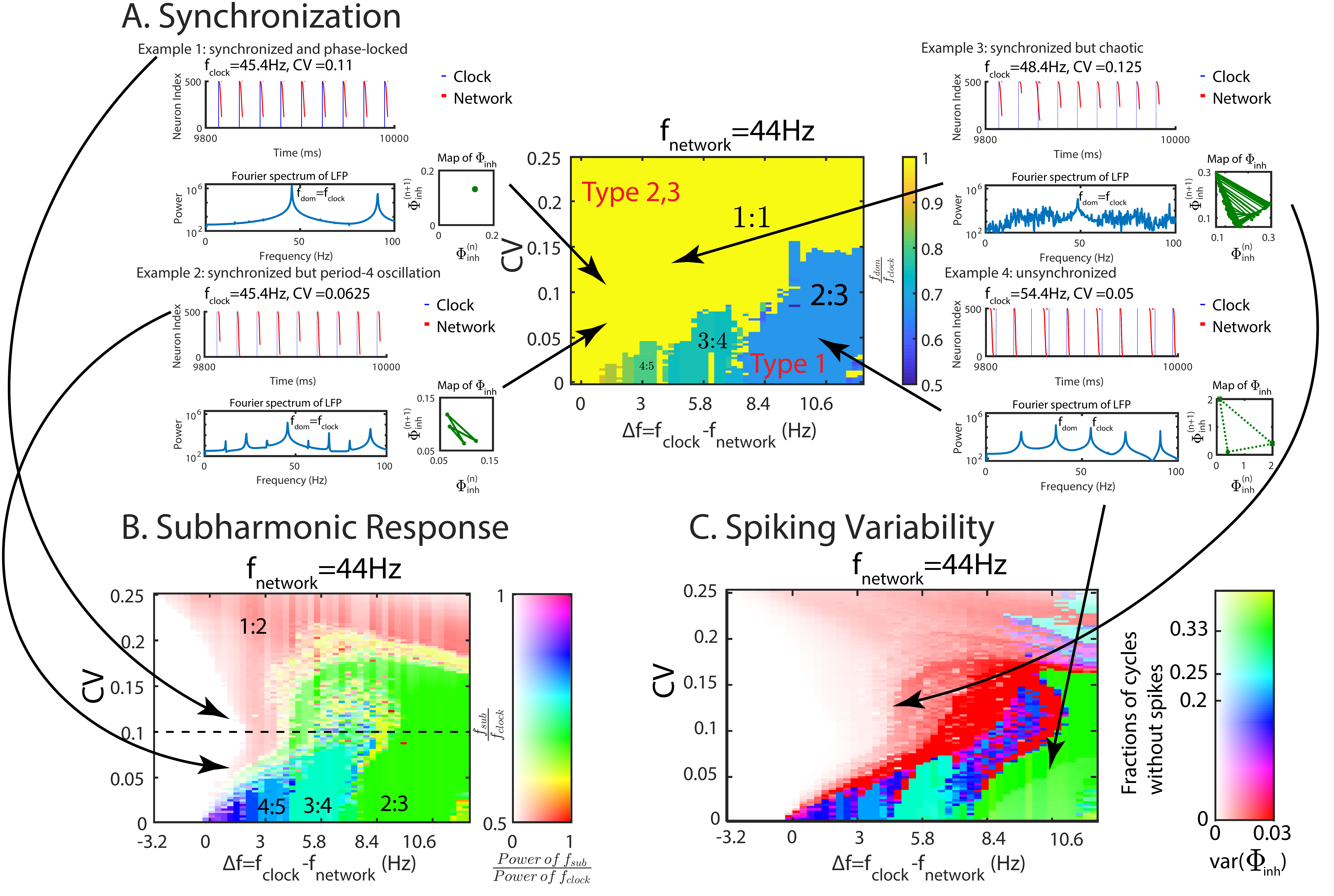}
\par\end{centering}
\caption{Network heterogeneity enhances synchronization and phase-locking of
periodically driven ING rhythm. (A) Synchronization quantified using
$f_{dom}:f_{clock}$ with $f_{dom}$ and $f_{clock}$ being the dominant
frequencies of the Fourier spectrum of the LFP of the network and
the clock, respectively. The neuronal heterogeneity enhanced the synchronization
by shifting $f_{dom}$ to $f_{clock}$. Example 1: Synchronized with
1:1 phase-locking. Example 2: Synchronized with subharmonic response
(period 4). Example 3: synchronized with subharmonic response (chaotic).
Example 4: Not synchronized. Squares and dashed lines in the map of
$\Phi_{inh}$ indicate clock cycles in which the network did not spike
($\Phi_{inh}$ was arbitrarily set to 2). (B) Subharmonic response.
Color hue and saturation indicate the frequency ratio $f_{sub}:f_{clock}$
and the ratio of the Fourier power at these two frequencies. $f_{sub}$
is the frequency of the dominant peak of the network power spectrum
that satisfies $f_{sub}<f_{clock}$. The power ratio is capped at
1. Dashed line marks the value of input heterogeneity used in Fig.\ref{fig:fmPRC-sketch}B.
(C) Spiking variability and $var(\Phi_{inh})$ as a function of neuronal
heterogeneity and detuning. Color hue indicates the fraction of clock
cycles without spikes in the network. In particular, red indicates
that the network spikes in every cycle. Color saturation indicates
$var(\Phi_{inh})$. The neuronal heterogeneity enhances the tightness
of the phase-locking. \label{fig:fmPRC-syn}}
\end{figure}

Having clarified the role of the fmPRC in the network's synchronizability
and ability to phase-lock, we investigated the role of neuronal heterogeneity
in more detail (Fig.\ref{fig:fmPRC-syn}). To do that, we adjusted
for each value of the input heterogeneity the mean input\textcolor{black}{{}
strength $I^{(I)}$} so as to keep the natural frequency $f_{network}$
constant ($f_{network}=44$ Hz). Then we determined the extent of
synchronization and phase-locking of the network under the influence
of periodic inhibitory input as a function of the detuning $\Delta f$
and network heterogeneity $CV$. As shown above, the fmPRC of a heterogeneous
network could be biphasic with the amplitude of the paradoxical phase
response increasing with neuronal heterogeneity. Expecting that for
sufficiently large heterogeneity an ING-rhythm could be accelerated
by a faster periodic inhibition, we tested phase-locking predominantly
for positive detuning, corresponding to $f_{clock}>f_{network}$.
We first investigated how neuronal heterogeneity affected the synchronization
by comparing the dominant frequency $f_{dom}$ in the Fourier spectrum
of the network's LFP with $f_{clock}$. In Fig.\ref{fig:fmPRC-syn}A,
the color hue indicates the ratio $f_{dom}:f_{clock}$. For small
heterogeneity, $f_{dom}$ was a rational multiple of $f_{clock}$
that depended on the detuning, while for sufficiently large CV the
network became synchronized in the sense that $f_{dom}=f_{clock}$
(yellow). The range of $\Delta f$ allowing synchronization became
wider with increasing neuronal heterogeneity, implying that the neuronal
heterogeneity enhanced the synchronization of the ING-rhythm. However,
note that $f_{dom}=f_{clock}$ did not imply a perfectly synchronized
or a 1:1 phase-locked state. In fact, various different subharmonic
responses arose: example 2 shows a period-4 state, while in example
3 the dynamics were actually chaotic (Fig.\ref{fig:fmPRC-syn}A) even
though $f_{dom}=f_{clock}$. Motivated by these observations, we divided
the states into three types:
\begin{itemize}
\item Type 1: $f_{dom}\neq f_{clock}$, not synchronized, not phase-locked
(Fig.\ref{fig:fmPRC-syn} example 4).
\item Type 2: $f_{dom}=f_{clock}$ with subharmonic response, might be poorly
phase-locked (Fig.\ref{fig:fmPRC-syn} example 3) or displaying rational
ratio phase-locking (Fig.\ref{fig:fmPRC-syn} example 2).
\item Type 3: $f_{dom}=f_{clock}$, no subharmonic response, 1-to-1 phase-locking
(Fig.\ref{fig:fmPRC-syn} example 1).
\end{itemize}
The phase diagram Fig.\ref{fig:fmPRC-syn}A does not differentiate
between types 2 and 3. It only shows that neuronal heterogeneity enhanced
the synchronization of the network by shifting $f_{dom}$ to $f_{clock}$.
Therefore, we studied whether neuronal heterogeneity also enhanced
the synchronization by weakening the subharmonic response and changing
the synchronized state from type 2 to type 3, as well as whether the
dynamics of the fmPRC shown in the bifurcation diagram Fig.\ref{fig:fmPRC-sketch}B
could predict the phase relationship between the network and the clock.
Using the same simulation setup as in Fig.\ref{fig:fmPRC-syn}A, the
subharmonic response is shown in Fig.\ref{fig:fmPRC-syn}B. The color
hue indicates the ratio $f_{sub}:f_{clock}$, where $f_{sub}$ is
the frequency of the dominant peak of the LFP power spectrum that
satisfies $f_{sub}<f_{clock}$. The color saturation gives the ratio
of the powers at $f_{sub}$ and $f_{clock}$ (capped at 1). Thus,
over most of the range of positive detuning and neuronal heterogeneity
tested, the fading-away of the color with increasing heterogeneity
reveals that the neuronal heterogeneity weakened the subharmonic response.
Over a small range of positive detuning, increasing neuronal heterogeneity
from small values induced perfect synchronization (type 3) by weakening
the subharmonic response with frequency ratio $f_{sub}:f_{clock}=1:2$;
the system traversed a continuous period-doubling bifurcation in reverse,
with type 2 (red) giving way to type 3 (white). Together with Fig.\ref{fig:fmPRC-syn}A,
this showed that neuronal heterogeneity could enhance the synchronization
both by making $f_{dom}=f_{clock}$ (from type 1 to type 2) and by
weakening the subharmonic response (from type 2 to type 3).\textbf{
}The range of detuning where increasing heterogeneity induced a type
3 synchronization became wider for larger synaptic delay within the
network (Suppl. Figure \ref{fig:fmPRC-syn-longerdelay})\textbf{.}
Note that the bifurcation diagram (Fig.\ref{fig:fmPRC-sketch}B) based
on the fmPRC agrees well with the subharmonic response marked along
the dashed line at $CV=0.1$ in Fig.\ref{fig:fmPRC-syn}B, suggesting
that the fmPRC can well predict the subharmonic response and persistent
phase response of the network.

In addition to enhancing the frequency synchonization, neuronal heterogeneity
was also able to increase the tightness of the phase-locking. Over
most of the parameter regime investigated, the variance of the phase
of the network relative to the periodic input ($var(\Phi_{inh})$)
decreased with increasing heterogeneity, as indicated by the decrease
in the color saturation in Fig.\ref{fig:fmPRC-syn}C. In fact, for
detuning between 0 Hz and 2 Hz the heterogeneity reduced $var(\Phi_{inh})$
to 0 (white), corresponding to the 1:1 phase-locked state. Even for
the 1:2 phase-locked state (cf. the red area in Fig.\ref{fig:fmPRC-syn}B)
$var(\Phi_{inh})$ was very small for a range of heterogeneity and
detuning (2 Hz to 4 Hz), indicating tight phase locking. Except for
type-3 synchronized states the size of the spike volleys varied between
clock cycles. In fact, over wide ranges of the parameters the network
did not spike in each of the clock cycles, as indicated by the color
hue in Fig.\ref{fig:fmPRC-syn}C, which gives the fraction of cycles
with no network spikes (e.g. Fig.\ref{fig:fmPRC-syn} example 4).

\subsubsection*{Paradoxical phase response and entrainment of PING rhythms}

Many $\gamma$-rhythms involve not only inhibitory neurons, but arise
from the mutual interaction of excitatory (E) and inhibitory (I) neurons
(PING rhythm) \cite{BaVi07}. The key elements to obtain a paradoxical
phase response and the ensuing enhanced synchronization are self-inhibition
within the network, neuronal heterogeneity and effective synaptic
delay. Since in PING rhythms the connections from E-cells to I-cells
and back to the E-cells form an effective self-inhibiting loop, we
asked whether PING-rhythms can exhibit behavior similar to the behavior
we identified for ING-rhythms.

\begin{figure}
\begin{centering}
\includegraphics[width=1\linewidth]{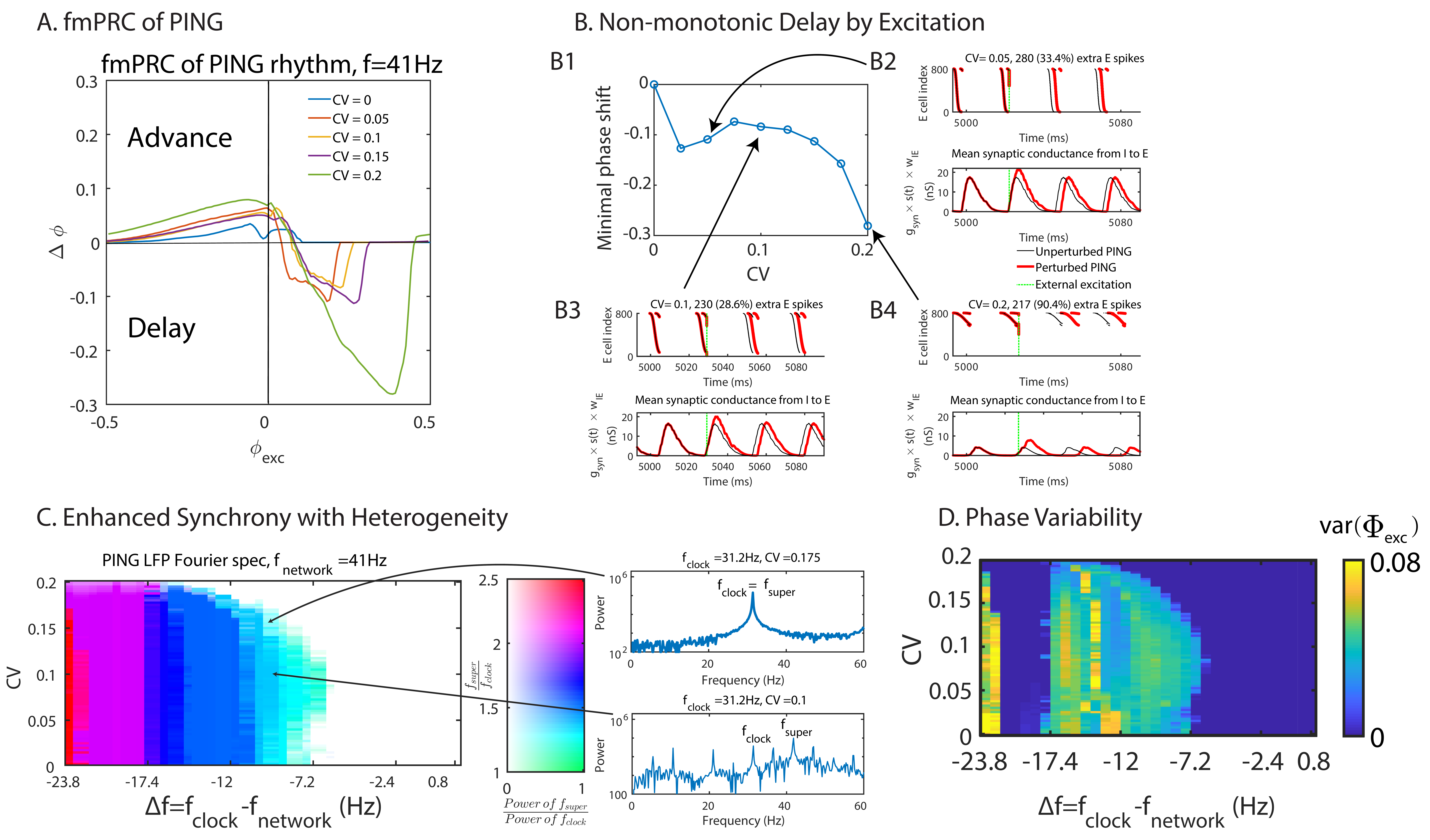}
\par\end{centering}
\caption{Network heterogeneity enhances the synchronization and the tightness
of phase-locking of the PING rhythm. (A) fmPRC of PING networks with
constant natural frequency ($f_{network}=41$ Hz) but different neuronal
heterogeneity. Only with neuronal heterogeneity the phase was delayed
by the excitation. (B) Non-monotonicity of the paradoxical delay with
constant natural frequency ($f_{network}=41$ Hz). B2-4: Top: raster
plot of spikes in E-population (input strength increased with cell
index). Bottom: mean inhibitory synaptic conductance within the PING
network. The titles show the absolute and relative increase in spike
number (B2: $CV=0.05$, B3: $CV=0.1$, B4: $CV=0.2$). (C) Subharmonic
response of the PING rhythm with periodic excitation as function of
neuronal heterogeneity and detuning. $f_{network}$ was fixed at 41
Hz. Color hue and saturation indicate the frequency ratio and power
ratio at the frequencies $f_{super}$ and $f_{clock}$ of the E-population's
LFP. $f_{super}$ was the frequency of the dominant peak of the LFP
power spectrum that satisfies $f_{super}>f_{clock}$. The power ratio
was capped at 1. Generally, the neuronal heterogeneity enhanced the
synchronization of the PING rhythm by weakening subharmonic response.
(D) The tightness of the phase-locking ($var(\Phi_{exc})$) as a function
of neuronal heterogeneity and detuning. The neuronal heterogeneity
enhanced the tightness of the phase-locking. For $\Delta f\in[-22\mbox{Hz},\,-17.4\mbox{Hz}]$
the clock was twice as fast as the network, resulting in vanishing
$var(\Phi_{exc})$. \label{fig:PING}}
\end{figure}

Considering a PING-rhythm generated by an E-I network comprised of
integrate-fire neurons, we first studied its fmPRC. To avoid that
all I-cells receive identical input and therefore spike as a single
unit, the I-cells received, in addition to the excitation from the
E-cells, heterogeneous, tonic, Gaussian-distributed subthreshold input
with mean $I^{(I)}=36$ pA and $CV^{(I)}$= 0.167. The phase response
of the network was probed by applying an identical external excitatory
perturbation to all E-cells and recording the resulting phase shift
of the LFP (cf. eqs.(\ref{eq:phi_exc},\ref{eq:phase_shift}))\textbf{
}of the E-population, averaged across 500 realizations of the subthreshold
input to the I-cells (Fig.\ref{fig:PING}A). More specifically, the
perturbations consisted of a square-wave excitatory current pulse
with amplitude 76 pA and duration 0.1 ms to each E-cell, resulting
in a 2 mV rapid depolarization. Without neuronal heterogeneity the
external excitation always advanced the phase of the rhythm resulting
in an fmPRC that was strictly positive. In the heterogeneous case,
however, the PING rhythm exhibited a paradoxical phase response, whereby
the collective rhythm was delayed while the individual neurons were
advanced by the excitation. The delay was caused by the increase of
self-inhibition within the network that was generated by the additional
spikes in the E-population, which in turn drove additional spikes
in the I-population. In contrast to the fmPRC of the ING-rhythm, this
paradoxical phase response was not monotonic in the heterogeneity.
While weak heterogeneity resulted in strong delay, the delay decreased
with increasing intermediate CV-values and only increased again for
larger CV (Fig.\ref{fig:PING}B left top). This non-monotonicity arose
because we kept the frequency of the network constant as we increased
its heterogeneity. This required a decrease in the tonic input to
the E-cells with increasing heterogeneity. For the stronger tonic
input used for weak heterogeneity the same external perturbation elicited
more additional spikes than it did for strong heterogeneity where
the tonic input was weaker (cf. titles of subpanels of Fig.\ref{fig:PING}B).
The total number of spikes occurring in each cycle of the unperturbed
network also decreased with increasing heterogeneity. Consequently,
the relative change in the number of spikes induced by the perturbation
was non-monotonic in the heterogeneity. As a result, the relative
change in the inhibitory synaptic conductance resulting from the perturbation
and with it the phase delay was also non-monotonic.

As for the ING rhythm, we investigated the role of neuronal heterogeneity
in the synchonizability and the ability of phase-locking of coupled
PING rhythms. In analogy to the ING-case, we considered the case of
the E-population of a PING network receiving periodic excitation generated
by a clock PING network (Fig.\ref{fig:Sketch}B). As before, we adjusted
the tonic input strength to the E-population to keep the natural frequency
of the network constant as we changed its heterogeneity ($f_{network}=41$Hz).
To probe the impact of the paradoxical phase response on the synchronization
we focused on negative detuning for which the periodic external excitation
needed to slow down the network in order to achieve phase-locking.
Indeed, with increasing heterogeneity the network could become synchronized
with the slower clock over a larger ranger of the detuning as indicated
by the fading saturation of the color in Fig.\ref{fig:PING}C. Here
the color hue indicates the ratio $f_{super}:f_{clock}$, where $f_{super}$
was determined as the frequency with the most power among the frequencies
higher than $f_{clock}$ in the Fourier spectrum of the E-population's
LFP. The color saturation indicates the ratio of the power at the
frequencies $f_{super}$ and $f_{clock}$. Thus, a color hue closer
to green ($f_{super}:f_{clock}=1:1$) or with a lower saturation implies
better synchronization. By observing how the width of the range of
detuning allowing synchronization varied with neuronal heterogeneity,
we concluded that generally, the synchronizability of PING rhythm
was enhanced by the neuronal heterogeneity by weakening subharmonic
response. Note that for $CV\in[0,0.1]$ the synchonizability of the
PING rhythm decreased slightly with neuronal heterogeneity. This was
consistent with the nonmonotonicity exhibited by the fmPRC seen in
Fig.\ref{fig:PING}B. The neuronal heterogeneity played a similar
role in the tightness of the phase-locking as in the synchronizability
(Fig.\ref{fig:PING}D).

\section{Discussion}

In this paper we have analyzed the response of collective oscillations
of inhibitory and of excitatory-inhibitory networks of integrate-fire
neurons to external perturbations. For ING- and PING-rhythms we have
shown that the combination of neuronal heterogeneity and effective
synaptic delay can qualitatively change the phase response compared
to the phase response of the individual neurons generating the rhythm.\textbf{
}Thus, perturbations that delay the I-cells can paradoxically advance
the ING-rhythm and perturbations that advance the E-cells can delay
the PING-rhythm. As a result, the macroscopic phase-response curve
for finite-amplitude perturbations (fmPRC) of the rhythm can change
sign as the phase of the perturbation is changed (type-II), even though
the PRC of all individual cells is strictly positive (type-I). This
change of the fmPRC enhances the ability of the $\gamma$-rhythm to
synchronize with other rhythms.

The key element of the mechanism driving the paradoxical phase response
and the enhanced synchronization is the cooperation of the external
perturbation and the effectively delayed within-network inhibition.
In the ING-network a suitably timed external perturbation delays the
lagging - but not the early - neurons sufficiently to allow the within-network
inhibition triggered by the early neurons to keep the lagging neurons
from spiking. This reduces the overall within-network inhibition and
with it the duration of the cycle. Thus, the perturbation modifies
the internal dynamics of the rhythm, which leads to changes in the
phase of the rhythm that can dominate the immediate phase change the
perturbation induces. The situation is somewhat similar to that investigated
in \cite{LePi10}. There it had been pointed out that an external
perturbation of a collective oscillation can lead to changes in its
phase in two stages: i) an immediate change of the phases of all oscillators
as a direct result of the perturbation and ii) a subsequent slower
change in the collective phase resulting from the convergence of the
disturbed phases back to the synchronized state. That analysis was
based on a network of phase oscillators and could therefore not include
a key element of our results, which is the perturbation-induced change
in the number of neurons that actually spike and the resulting change
in the within-network inhibition that results in a change of the period
of the rhythm. As discussed in \cite{SerKop13,CanKop14}, for ING-rhythms
such a change in the number of spiking neurons underlies also the
enhanced phase-locking found in \cite{WhBa00}.

Going beyond ING-rhythms, we showed that PING-rhythms can also exibit
a paradoxical phase response \emph{via }a mechanism that is analogous
to that of ING-rhythms. For that analysis we have focused on excitatory-inhibitory
networks with only connections between but not within the excitatory
and inhibitory populations. For excitatory inputs to the excitatory
cells to generate a paradoxical phase response it is necessary that
the additional spikes of the excitatory neurons that are caused by
the external perturbation induce additional spikes of the inhibitory
neurons. This behavior arises if the inibitory population is also
allowed to be heterogeneous. Moreover, the within-network inhibition
has to be strong enough to be able to suppress the spiking of lagging
excitatory neurons. This is, e.g., found in mice piriform cortex,
where principal neurons driven by sensory input from the olfactory
bulb arriving early during a sniff recruit inhibitory interneurons
via long-range recurrent connections, resulting in the global, transient
suppression of subsequent cortical activity \cite{BolFra18}. A characteristic
feature of the paradoxical phase response of the PING rhythm is the
extended cycle time following enhanced activation of the excitatory
cells. A strong such correlation between the cycle time and the previous
LFP amplitude has been observed for the $\gamma$-rhythm in hippocampus
\cite{AtSc09}. To assess whether this rhythm exhibits paradoxical
phase response would require comparing the macroscopic phase response
\cite{AkaKul12a} with that of indvidual participating neurons.

In order for the global perturbation to affect the various neurons
differently, they have to be at different phases in their cycle. 
Our analysis suggests that the specific cause for this heterogeneity
in the spike times does not play an important role. Indeed, as shown
in \cite{MenRie18}, even fluctuating heterogeneities that are generated
by noise rather than static heterogeneities reflecting intrinsic properties
of neurons can enhance the synchronization of multiple $\gamma$-rhythms
in interconnected networks of identical neurons. Note that the noise
driving this synchronization is uncorrelated across neurons. The analysis
of that system revealed the same mechanism at work as the one identified
here.\textcolor{red}{}

In various previous analytical and computational analyses it has been
found that the dynamics of the macroscopic phase of a collective oscillation
can qualitatively differ from that of the microscopic phase. Thus,
for interacting groups of noisy identical phase oscillators the macroscopic
phases of the groups can tend to lign up with each other, even if
all pair-wise interactions between individual oscillators prefer the
antiphase state, and vice versa \cite{KaNa10}. An analogous result
has been obtained for heterogeneous populations of noiseless oscillators
\cite{KaNa10a}.

Qualitative changes have also been found in the macroscopic phase
response of rhythms in noisy homogeneous networks when the noise level
was changed \cite{KawKur11,KotErm14,AkaKot18}. Using a Fokker-Planck
approach for globally coupled excitable neurons, a type-I mPRC was
obtained for weak noise, where the rhythm emerges through a SNIC bifurcation,
while a type-II mPRC arose for strong noise that led to a Hopf bifurcation
\cite{KawKur11}. A similar approach was used to obtain the mPRC via
the adjoint method for an extension of the theta-model that implements
conductance-based synaptic interactions. Again, although individual
theta-neurons have a type-I PRC, a type-II mPRC was obtained for the
rhythm, which arose in a Hopf bifurcation \cite{KotErm14}. This was
also the case in an extension to networks of excitable and inhibitory
neurons \cite{AkaKot18}.

Thus, results reminiscent of those presented here have been obtained
previously. However, the mechanism underlying them was not addressed
in detail and remained poorly understood. We expect that our analysis
will provide insight into those systems. The key element of the mechanism
discussed here is that due to the dispersion of the spike times, which
either results from neuronal heterogeneity or noise, the external
perturbation enables the within-network inhibition to suppress the
spiking of a larger number of neurons than without it. In our system
this was facilitated by the delay with which spikes triggered the
within-network inhibition, which allowed some neurons to escape its
impact in the absence of the external perturbation, but not in its
presence. Our analysis showed, however, that the explicit delay is
not necessary; the effective delay resulting from a double-exponential
synaptic interaction is sufficient. In fact, when reducing that effective
delay the paradoxical phase response did not disappear until the delay
was so short that the rhythm itself developed a strong subharmonic
component and disintegrated.

In this paper we have focused on a specific, very simple neuronal
model, the leaky integrate-fire model with conductance-based pulsatile
coupling. In previous work on the enhanced synchronization among $\gamma$-rhythms
\emph{via} noise-induced spiking heterogeneity it was demonstrated
that this result does not depend sensitively on the neuron type. Comparable
results were obtained also with Morris-Lecar neurons for parameters
in which the periodic spiking arises from a SNIC-bifurcation, resulting
in a type-I PRC as is the case for integrate-fire neurons, but also
for parameters for which the spiking is due to a Hopf bifurcation,
resulting in a type-II PRC \cite{MenRie18}. For networks comprised
of heterogeneous neurons with type-II PRC the fmPRC of the collective
oscillation is likely to be more complex, since the heterogeneity
allows the same input to induce phase shifts with opposite signs for
different neurons. However, we expect that the interplay between the
within-network inhibition and the external perturbation can again
substantially and qualitatively modify the fmPRC by changing the number
of neurons participating in the rhythm.

In \cite{MenRie18} the results were also found to be robust with
respect to significant changes in the network connectivity (random
instead of all-to-all) as well as the reversal potential of the inhibitory
synapses, as long as the rhythm itself persisted robustly (cf. \cite{BoKr10}).
In fact, the coupling did not even have to be synaptic; collective
oscillations of relaxation-type chemical oscillators that were coupled
diffusively were also shown to exhibit noise-induced synchronization.
These results suggest that the paradoxical phase response found here
arises in a much wider class of macroscopic collective oscillations.

The strong paradoxical phase response that we demonstrated for heterogeneous
networks allows their rhythm to synchronize with a periodic external
input over a range of detuning that increases substantially with the
neuronal heterogeneity. This is reminiscent of computational results
for anterior cingulate cortex that investigated networks of excitatory
neurons coupled via a common population of inhibitory neurons. There
heterogeneity was also found to enhance the synchrony of rhythms,
as measured in terms of coincident spikes within 10ms bins \cite{AdaLeB17}.

The heterogeneity-enhanced synchrony we have identified suggests that
the coherence of $\gamma$-rhythms emerging in different interacting
networks could also be enhanced by neuronal heterogeneity. It has
been proposed that the coherence of different $\gamma$-rhythms, which
has been observed to be modified by attention \cite{BosFri12}, plays
an important role in the communication between the corresponding networks
\cite{Fr05,Fri15}. Computational studies have shown that the direction
of information transfer between networks depends on the relative phase
of their rhythms \cite{DumGut19,PalBat17}, which can be changed by
switching between different base states \cite{WitBat13,KirBat16}.
Whether the enhanced synchrony resulting from neuronal heterogeneity
enhances this information transfer is still an open question.

Disrupted $\gamma$-rhythms have been observed in multiple brain regions
in neurological diseases, especially Alzheimer\textquoteright s disease.
Optogenetic and sensory periodic stimulation at $\gamma$-frequencies
has been found to entrain the $\gamma$-rhythm in hippocampus and
visual cortex, respectively, and has resulted in a significant reduction
in total amyloid level \cite{IacTsa16}. Similar neuro-protective
effects of entrainment by external $\gamma$-stimulation have also
been found for other sensory modalities \cite{MarTsa19,AdaTsa19}.
This suggests that $\gamma$-stimulation by sensory input might be
a feasible therapeutic approach. Our results suggest a potential role
of neuronal heterogeneity in this context.

From a functional perspective, it has been shown that the noise-induced
synchronization mentioned above can facilitate certain learning processes
\cite{MenRie18a}. Specifically, a read-out neuron was considered
that received input from neurons in two networks \emph{via} synapses
that exhibited spike-timing dependent plasticity. The two networks
were interacting with each other and each of them exhibited a $\gamma$-rhythm,
albeit at different frequencies. For low noise the two rhythms were
not synchronized and the read-out neuron received inputs from the
two networks at uncorrelated times. These inputs drove the plasticity
inconsistently, leading only to a very slow overall evolution of the
synaptic weights, if any. However, for stronger noise the two networks
were synchronized, providing a more consistent spike timing that lead
to substantial changes in the synaptic weights. As a result, the read-out
neuron was eventually only driven by the network that had the larger
natural frequency in the absence of the coupling between the networks.
It is expected that synchrony by neuronal heterogeneity will have
a similar impact.

\section{Methods}

\textbf{Neuron model.} Both E-cells and I-cells were modeled as leaky
integrate-and-fire neurons, each characterized by a membrane potential
$V_{i}(t)$ satisfying 
\begin{equation}
\tau_{E,I}\frac{d}{dt}V_{i}=-(V_{i}-V_{rest})+\frac{I_{i}^{(syn)}}{g_{syn}}+\frac{I_{i}^{(ext)}}{g_{ext}}+\frac{I_{i}^{(bias)}}{g_{bias}}\,,
\end{equation}

where $V_{rest}$ is the resting potential and $\tau_{E,I}$ the membrane
time constants of the E- and I-cells, respectively. $I_{i}^{(syn)}(t)$
is the total synaptic current that the neuron receives from the other
neurons within the network. $I_{i}^{(ext)}(t)$ is a time-dependent
external input that represents perturbations applied to determine
the fmPRC or, in the study of synchronization, the periodic input
generated by the clock network. $I_{i}^{(bias)}$ denotes a tonic,
neuron-specific excitatory bias current that implements the heterogeneity
of the neuron properties. The corresponding conductances are denoted
$g_{syn}$, $,g_{ext}$, and $g_{bias}$. Upon the $i^{\mbox{th}}$
neuron reaching the spiking threshold $V_{peak}$, the voltage $V_{i}$
was reset to the fixed value $V_{reset}$. Parameters for the neuron
were kept fixed throughout all simulations (see Table \ref{tab:neuron-model}).
The local field potential (LFP) of the network was approximated as
the mean voltage across all neurons $j=1,...N$ in the respective
population.

\textbf{Network model. }We studied two types of networks: an ING network
and a PING network. The ING network was modeled as an all-to-all inhibitory
network of $N_{I}^{(ING)}$ interneurons. The PING network was modeled
as a network of $N_{I}^{(PING)}$ interneurons and $N_{E}^{(PING)}$
principal cells with all-to-all interneuron-principal and principal-interneuron
connections (i.e., without principal-principal and interneuron-interneuron
connections). In PING, only principal cells received external input
$I^{ext}(t)$.

To gain insight into the interaction between two ING rhythms, we considered
the simplified situation in which all neurons in the network received
strictly periodic input $I^{(ext)}$, which was generated by another
ING network (`clock'). Similarly, for PING rhythms, the E-cells of
the PING network received strictly periodic excitatory input $I^{(ext)}$
from another PING network through all-to-all connection between their
E populations.

\textbf{Synaptic currents.} We used delayed double-exponential conductance-based
currents to model the excitatory and the inhibitory synaptic inputs
from neuron $j$ to neuron $i$,

\begin{equation}
I_{ij}^{(syn)}(t)=g_{syn}\frac{\tau_{E,I}}{\tau_{2}^{E,I}-\tau_{1}^{E,I}}\left(A_{ij}^{(2)}(t)-A_{ij}^{(1)}(t)\right)\,\left(V_{rev,j}-V_{i}(t)\right)\,,\label{eq:double_exp_1}
\end{equation}

with the two exponentials $A_{ij}^{(1,2)}(t)$ satisfying

\begin{equation}
\frac{d}{dt}A_{ij}^{(1,2)}(t)=-\frac{A_{ij}^{(1,2)}(t)}{\tau_{1,2}^{E,I}}+\sum_{k}W_{ij}\delta(t-t_{j}^{(k)}-\tau_{d})\,,\label{eq:double_exp_2}
\end{equation}

where $V_{rev,j}$ is the synaptic reversal potential corresponding
to the synapse type, $W_{ij}$ the dimensionless synaptic strength,
and $\delta$ the Dirac $\delta$-function. All synapses of the same
type (I-I, I-E, E-I) were equally strong. The time constants of $A_{i}^{(1,2)}(t)$
satisfied $\tau_{2}^{E,I}>\tau_{1}^{E,I}$. The synaptic current was
normalized to render the time integral independent of the synaptic
time constants $\tau_{1,2}^{E,I}$. The inhibitory synaptic currents
had a slower decay than the excitatory ones (cf. Table \ref{tab:neuron-model}).
We included an explicit synaptic delay $\tau_{d}$ in the model\textbf{.}
Every spike of the presynaptic neuron $j$ at time $t_{j}^{(k)}$
triggered a jump in both $A_{ij}^{(1,2)}(t)$, making the synaptic
conductance rise continously after a synaptic delay $\tau_{d}$.

External periodic inputs were also modeled as double-exponential conductance-based
currents with $g_{syn}$ in (\ref{eq:double_exp_1},\ref{eq:double_exp_2})
replaced by \textbf{$g_{ext}$}. The time constants and delay were
as for the within-network synaptic inputs.

\textbf{Heterogeneous tonic input.} The bias currents $I_{i}^{(bias)}$
of the ING network were Gaussian distributed around $I_{mean}$ with
a coefficient of variation $CV$ and arranged in increasing order,
$I_{1}^{(bias)}>I_{2}^{(bias)}...>I_{N}^{(bias)}$. For the PING network,
all excitatory neurons received a heterogeneous bias $I_{E}^{(bias)}$
with mean $I^{(E)}$ and a coefficient of variation $CV^{(E)}$. Similarly,
the bias currents $I_{I}^{(bias)}$ to the inhibitory neurons were
characterized by their mean $I^{(I)}$and their coefficient of variation
$CV^{(I)}$. Without the excitatory input from principal cells, the
voltage of interneurons remained below the spiking threshold. In our
investigation of the impact of the neuronal heterogeneity on the phase
response and entrainment of the PING rhythm we kept $CV^{(I)}$ fixed
and varied $CV^{(E)}$.

\textbf{Macroscopic Phase-response Curve for Finite-Amplitude Perturbations
(fmPRC).}

\textbf{ING rhythm. }For a single ING network, we applied a single
inhibitory $\delta$-pulse to each neuron $j=1,...N_{I}^{(ING)}$
at time $t_{inh}$ (dashed green line in Fig.\ref{fig:Sketch-of-model}B)
and recorded the resulting phase shift $\Delta\varphi$. The amplitude
of the inhibitory perturbation to each neuron was the same. The phase
of the inhibition was defined as

\begin{equation}
\phi_{inh}=\frac{t_{inh}-t_{firstspike}^{(unperturbed)}}{T},\label{eq:phi_inh_def}
\end{equation}

where $T$ was the period of the network LFP and $t_{firstspike}^{(unperturbed)}$
the time of the first spike in the spike volley of the unperturbed
network that was closest to $t_{inh}$. The resulting phase shift
$\Delta\phi$ was given by

\begin{equation}
\Delta\phi=\frac{\left(t_{firstspike}^{(unperturbed)}-t_{firstspike}^{(perturbed)}\right)}{T},
\end{equation}

where $t_{firstspike}^{(perturbed)}$ is the time of the first spike
in the corresponding volley in the perturbed network. $\Delta\phi$
and $\phi_{inh}$ were taken to be in the range $[-0.5\,\:0.5)$.
Positive $\Delta\phi$ indicated that the network was advanced by
the perturbation, while negative indicated a delay.

The periodic input (`clock') that was used to test the synchronizability
of the ING-rhythm was generated by a homogeneous ING network. The
phase of the periodic input in the $n^{\mbox{th}}$ clock cycle was
defined by

\begin{equation}
\Phi_{inh}^{(n)}=\frac{\left(t_{firstspike}^{(clock)(n)}+\tau_{d}-t_{firstspike}^{(network)(n)}\right)}{T},\label{eq:Phi_inh_def}
\end{equation}

where $t_{firstspike}^{(network)(n)}$ was the time of the first spike
in the spike volley of the network in the $n^{\mbox{th}}$ cycle and
$t_{firstspike}^{(clock)(n)}$ the time of the spike of the clock.
In contrast to the definition of $\phi_{inh}$ in (\ref{eq:phi_inh_def}),
the definition of $\Phi_{inh}^{(n)}$ included the delay $\tau_{d}$,
since the inhibition generated by the clock arrived with delay $\tau_{d}$
in the network.

\textbf{PING rhythm. }To probe the phase response of the PING network
we used the same approach as for the ING rhythm, except that we used
excitatory instead of inhibitory $\delta$-pulses and applied them
only to the E-cells. The phase of the excitation $\phi_{exc}$ and
the resulting phase shift $\Delta\phi$ were determined similarly
as in the case of the ING rhythm,

\begin{equation}
\phi_{exc}=\frac{t_{exc}-t_{firstspike}^{(unperturbed)}}{T},\label{eq:phi_exc}
\end{equation}

\begin{equation}
\Delta\phi=\frac{(t_{firstspike}^{(unperturbed)}-t_{firstspike}^{(perturbed)})}{T},\label{eq:phase_shift}
\end{equation}

where $t_{firstspike}^{(perturbed)}$ and $t_{firstspike}^{(unperturbed)}$
were the times of the first spike in the respective spike volleys
of the E-population.

Analogously to $\Phi_{inh}^{(n)}$, the phase of the periodic input
in the $n^{\mbox{th}}$ clock cycle was given by

\begin{equation}
\Phi_{exc}^{(n)}=\frac{(t_{firstspike}^{(clock)(n)}+\tau_{d}-t_{firstspike}^{(network)(n)})}{T}.
\end{equation}
Throughout, the tonic, Gaussian distributed input to the interneurons
in the PING network was fixed: $I^{(I)}=36$ pA, $CV^{(I)}=0.167$.

\begin{flushleft}
\begin{table}[H]
\begin{raggedright}
\begin{tabular*}{5cm}{@{\extracolsep{\fill}}>{\raggedright}p{5cm}>{\centering}p{2cm}}
\textbf{ING network} & \tabularnewline
Parameter & Value\tabularnewline
$\tau_{I}$, membrane time constant & 20 ms\tabularnewline
$u_{rest}$, resting potential & -55 mV\tabularnewline
$V_{peak}$, spiking threshold & -50 mV\tabularnewline
$V_{reset}$, reset voltage & -60 mV\tabularnewline
$\tau_{d}$, synaptic delay & 3 ms\tabularnewline
$N_{I}^{(ING)}$, \# of interneurons & 500\tabularnewline
$W$, synaptic strength within the network & $7.5\times10^{-3}$\tabularnewline
$W^{(ext)}$, synaptic strength for the input from the clock network & $1.8\times10^{-3}$\tabularnewline
\end{tabular*} $\qquad\qquad\qquad\qquad\qquad$%
\begin{tabular*}{2cm}{@{\extracolsep{\fill}}>{\raggedright}p{7cm}>{\raggedright}p{2cm}}
\textbf{Synaptic currents} & \tabularnewline
Parameter & Value\tabularnewline
$\tau_{1}^{E}$, time constant of rise in excitatory synapse & 0.5 ms\tabularnewline
$\tau_{2}^{E}$, time constant of decay in excitatory synapse & 2 ms\tabularnewline
$\tau_{1}^{I}$, time constant of rise in inhibitory synapse & 0.5 ms\tabularnewline
$\tau_{2}^{I}$, time constant of decay in inhibitory synapse & 5 ms\tabularnewline
$V_{rev}^{I}$, reversal potential of inhibitory synapse & -70 mV\tabularnewline
$V_{rev}^{E}$, reversal potential of excitatory synapse & 0 mV\tabularnewline
\end{tabular*}
\par\end{raggedright}
\raggedright{}%
\begin{tabular*}{2cm}{@{\extracolsep{\fill}}>{\raggedright}p{5cm}>{\raggedright}p{2cm}}
\textbf{PING network} & \tabularnewline
Parameter & Value\tabularnewline
$\tau_{E}$, membrane time constant of principal cells & 20 ms\tabularnewline
$\tau_{I}$, membrane time constant of interneurons & 10 ms\tabularnewline
$u_{rest}$, resting potential & -70 mV\tabularnewline
$V_{peak}$, spiking threshold & -52 mV\tabularnewline
$V_{reset}$, reset voltage & -59 mV\tabularnewline
$\tau_{d}$, synaptic delay & 1 ms\tabularnewline
$N_{I}^{(PING)}$, \# of interneurons & 200\tabularnewline
$N_{E}^{(PING)}$, \# of principal cells & 800\tabularnewline
$W^{I}$, inhibitory synaptic strength within the network & $5.4\times10^{-3}$\tabularnewline
$W^{E}$, excitatory synaptic strength within the network & $1.67\times10^{-3}$\tabularnewline
$W^{(ext)}$, clock-network synaptic strength & $1.6\times10^{-3}$\tabularnewline
\end{tabular*}$\qquad\qquad\,\,\,\,\qquad\qquad$$\qquad\qquad\qquad\qquad\qquad$%
\begin{tabular*}{2cm}{@{\extracolsep{\fill}}>{\raggedright}p{7cm}>{\raggedright}p{2cm}}
\textbf{Synaptic conductances} & \tabularnewline
Parameter & Value\tabularnewline
Excitatory input on principal cells : $g_{ext}^{(PING)}$, $g_{bias\:E}^{(PING)}$ & 0.19 nS\tabularnewline
Excitatory input on interneurons: $g_{bias}^{(ING)}$, $g_{syn\:EtoI}^{(PING)}$,
$g_{bias\:I}^{(PING)}$ & 0.3 nS\tabularnewline
Inhibitory input on principal cells: $g_{syn\:ItoE}^{(PING)}$ & 2.5 nS\tabularnewline
Inhibitory input on interneurons: $g_{ext}^{(ING)}$, $g_{syn}^{(ING)}$ & 4 nS\tabularnewline
\end{tabular*}\caption{\textbf{Parameters used in the model.} Most parameters are based on
\cite{MenRie18,BrWa03}. \label{tab:neuron-model}}
\end{table}
\par\end{flushleft}

\newpage

\bibliographystyle{plos2015}
\bibliography{journal}

\begin{thebibliography}{10}

\bibitem{BrJo05a}
Bruesselbach H, Jones DC, Mangir MS, Minden M, Rogers JL.
\newblock Self-organized coherence in fiber laser arrays.
\newblock Optics Letters. 2005;30(11):1339--1341.
\newblock doi:{10.1364/OL.30.001339}.

\bibitem{WiCo96}
Wiesenfeld K, Colet P, Strogatz SH.
\newblock Synchronization Transitions in a Disordered {Josephson} Series Array.
\newblock Phys Rev Lett. 1996;76(3):404--407.

\bibitem{LiWe97}
Liu C, Weaver DR, Strogatz SH, Reppert SM.
\newblock Cellular Construction of a Circadian Clock: Period Determination in
  the Suprachiasmatic Nuclei.
\newblock Cell. 1997;91(6):855--860.

\bibitem{VenOat20}
Venzin OF, Oates AC.
\newblock What are you synching about? Emerging complexity of Notch signaling
  in the segmentation clock.
\newblock Developmental Biology. 2020;460(1):40--54.
\newblock doi:{10.1016/j.ydbio.2019.06.024}.

\bibitem{Wa10}
Wang XJ.
\newblock Neurophysiological and Computational Principles of Cortical Rhythms
  in Cognition.
\newblock Physiol Rev. 2010;90(3):1195--1268.

\bibitem{BoEp05}
{B{\"o}rgers} C, Epstein S, Kopell NJ.
\newblock Background gamma rhythmicity and attention in cortical local
  circuits: a computational study.
\newblock Proc Natl Acad Sci U S A. 2005;102(19):7002--7007.

\bibitem{BoKo08}
B{\"o}rgers C, Kopell NJ.
\newblock Gamma oscillations and stimulus selection.
\newblock Neural Comput. 2008;20(2):383--414.

\bibitem{BosFri12}
Bosman CA, Schoffelen JM, Brunet N, Oostenveld R, Bastos AM, Womelsdorf T,
  et~al.
\newblock Attentional stimulus selection through selective synchronization
  between monkey visual areas.
\newblock Neuron. 2012;75:875--888.
\newblock doi:{10.1016/j.neuron.2012.06.037}.

\bibitem{RobDeW13}
Roberts MJ, Lowet E, Brunet NM, Ter~Wal M, Tiesinga P, Fries P, et~al.
\newblock Robust Gamma Coherence between Macaque {V1} and {V2} by Dynamic
  Frequency Matching.
\newblock Neuron. 2013;78(3):523--536.
\newblock doi:{10.1016/j.neuron.2013.03.003}.

\bibitem{BuzSch15}
Buzsaki G, Schomburg EW.
\newblock What does gamma coherence tell us about inter-regional neural
  communication?
\newblock Nature Neuroscience. 2015;18(4):484--489.
\newblock doi:{10.1038/nn.3952}.

\bibitem{Fri15}
Fries P.
\newblock Rhythms for Cognition: Communication through Coherence.
\newblock Neuron. 2015;88:220--235.
\newblock doi:{10.1016/j.neuron.2015.09.034}.

\bibitem{PalBat17}
Palmigiano A, Geisel T, Wolf F, Battaglia D.
\newblock Flexible information routing by transient synchrony.
\newblock Nature Neuroscience. 2017;20:1014--1022.
\newblock doi:{10.1038/nn.4569}.

\bibitem{DumGut19}
Dumont G, Gutkin B.
\newblock Macroscopic phase resetting-curves determine oscillatory coherence
  and signal transfer in inter-coupled neural circuits.
\newblock PLoS Computational Biology. 2019;15:e1007019.
\newblock doi:{10.1371/journal.pcbi.1007019}.

\bibitem{AdaTsa19}
Adaikkan C, Middleton SJ, Marco A, Pao PC, Mathys H, Kim DNW, et~al.
\newblock Gamma Entrainment Binds Higher-Order Brain Regions and Offers
  Neuroprotection.
\newblock Neuron. 2019;102:929--943.e8.
\newblock doi:{10.1016/j.neuron.2019.04.011}.

\bibitem{SchBut12}
Schultheiss NW, Butera RJ.
\newblock Phase Response Curves in Neuroscience.
\newblock Springer; 2012.

\bibitem{BrMo04}
Brown E, Moehlis J, Holmes P.
\newblock On the Phase Reduction and Response Dynamics of Neural Oscillator
  Populations.
\newblock Neural Comput. 2004;16(4):673--715.

\bibitem{KaNa10a}
Kawamura Y, Nakao H, Arai K, Kori H, Kuramoto Y.
\newblock Phase synchronization between collective rhythms of globally coupled
  oscillator groups: noiseless nonidentical case.
\newblock Chaos (Woodbury, NY). 2010;20:043110.
\newblock doi:{10.1063/1.3491346}.

\bibitem{LePi10}
Levnajic Z, Pikovsky A.
\newblock Phase resetting of collective rhythm in ensembles of oscillators.
\newblock Physical review E, Statistical, nonlinear, and soft matter physics.
  2010;82:056202.
\newblock doi:{10.1103/PhysRevE.82.056202}.

\bibitem{KaNa08}
Kawamura Y, Nakao H, Arai K, Kori H, Kuramoto Y.
\newblock Collective Phase Sensitivity.
\newblock Phys Rev Lett. 2008;101(2):024101.

\bibitem{KaNa10}
Kawamura Y, Nakao H, Arai K, Kori H, Kuramoto Y.
\newblock Phase Synchronization Between Collective Rhythms of Globally Coupled
  Oscillator Groups: Noisy Identical Case.
\newblock Chaos. 2010;20(4):043109.

\bibitem{KawKur11}
Kawamura Y, Nakao H, Kuramoto Y.
\newblock Collective phase description of globally coupled excitable elements.
\newblock Physical review E, Statistical, nonlinear, and soft matter physics.
  2011;84:046211.
\newblock doi:{10.1103/PhysRevE.84.046211}.

\bibitem{KotErm14}
Kotani K, Yamaguchi I, Yoshida L, Jimbo Y, Ermentrout GB.
\newblock Population dynamics of the modified theta model: macroscopic phase
  reduction and bifurcation analysis link microscopic neuronal interactions to
  macroscopic gamma oscillation.
\newblock J R Soc Interface. 2014;11(95):20140058.
\newblock doi:{10.1098/rsif.2014.0058}.

\bibitem{MonRox15}
Montbrio E, Pazo D, Roxin A.
\newblock Macroscopic description for networks of spiking neurons.
\newblock Physical Review X. 2015;5:021028.

\bibitem{OtAn08}
Ott E, Antonsen TM.
\newblock Low dimensional behavior of large systems of globally coupled
  oscillators.
\newblock Chaos. 2008;18(3):037113.
\newblock doi:{10.1063/1.2930766}.

\bibitem{LukSo13}
Luke TB, Barreto E, So P.
\newblock Complete Classification of the Macroscopic Behavior of a
  Heterogeneous Network of Theta Neurons.
\newblock Neural Computation. 2013;25(12):3207--3234.
\newblock doi:{10.1162/NECO\_a\_00525}.

\bibitem{HanFor15}
Hannay KM, Booth V, Forger DB.
\newblock Collective phase response curves for heterogeneous coupled
  oscillators.
\newblock Physical Review E. 2015;92(2):022923.
\newblock doi:{10.1103/PhysRevE.92.022923}.

\bibitem{DumGut17}
Dumont G, Ermentrout GB, Gutkin B.
\newblock Macroscopic phase-resetting curves for spiking neural networks.
\newblock Physical Review E. 2017;96:042311.
\newblock doi:{10.1103/PhysRevE.96.042311}.

\bibitem{AkaKot18}
Akao A, Ogawa Y, Jimbo Y, Ermentrout GB, Kotani K.
\newblock Relationship between the mechanisms of gamma rhythm generation and
  the magnitude of the macroscopic phase response function in a population of
  excitatory and inhibitory modified quadratic integrate-and-fire neurons.
\newblock Physical Review E. 2018;97(1):012209.
\newblock doi:{10.1103/PhysRevE.97.012209}.

\bibitem{MenRie18}
Meng JH, Riecke H.
\newblock Synchronization by uncorrelated noise: interacting rhythms in
  interconnected oscillator networks.
\newblock Scientific Reports. 2018;8:6949.
\newblock doi:{10.1038/s41598-018-24670-y}.

\bibitem{WhBa00}
White JA, Banks MI, Pearce RA, Kopell NJ.
\newblock Networks of interneurons with fast and slow gamma-aminobutyric acid
  type A (GABA(A)) kinetics provide substrate for mixed gamma-theta rhythm.
\newblock Proc Nat Acad Sci USA. 2000;97(14):8128--8133.
\newblock doi:{10.1073/pnas.100124097}.

\bibitem{SerKop13}
Serenevy AK, Kopell NJ.
\newblock Effects of Heterogeneous Periodic Forcing on Inhibitory Networks.
\newblock Siam Journal on Applied Dynamical Systems. 2013;12(3):1649--1684.
\newblock doi:{10.1137/12089274X}.

\bibitem{Fr05}
Fries P.
\newblock A Mechanism for Cognitive Dynamics: Neuronal Communication Through
  Neuronal Coherence.
\newblock Trends Cogn Sci. 2005;9(10):474--480.

\bibitem{BaVi07}
Bartos M, Vida I, Jonas P.
\newblock Synaptic mechanisms of synchronized gamma oscillations in inhibitory
  interneuron networks.
\newblock Nature Reviews Neuroscience. 2007;8(1):45--56.
\newblock doi:{10.1038/nrn2044}.

\bibitem{CanKop14}
Cannon J, McCarthy MM, Lee S, Lee J, B\"orgers C, Whittington MA, et~al.
\newblock Neurosystems: brain rhythms and cognitive processing.
\newblock European J Neuroscience. 2014;39:705--719.
\newblock doi:{10.1111/ejn.12453}.

\bibitem{BolFra18}
Bolding KA, Franks KM.
\newblock Recurrent cortical circuits implement concentration-invariant odor
  coding.
\newblock Science (New York, NY). 2018;361.
\newblock doi:{10.1126/science.aat6904}.

\bibitem{AtSc09}
Atallah BV, Scanziani M.
\newblock Instantaneous Modulation of Gamma Oscillation Frequency by Balancing
  Excitation with Inhibition.
\newblock Neuron. 2009;62(4):566--577.
\newblock doi:{10.1016/j.neuron.2009.04.027}.

\bibitem{AkaKul12a}
Akam T, Oren I, Mantoan L, Ferenczi E, Kullmann DM.
\newblock Oscillatory dynamics in the hippocampus support dentate gyrus-CA3
  coupling.
\newblock Nature Neuroscience. 2012;15(5):763--768.
\newblock doi:{10.1038/nn.3081}.

\bibitem{BoKr10}
B{\"o}rgers C, Krupa M, Gielen S.
\newblock The Response of a Classical {Hodgkin}-{Huxley} Neuron To an
  Inhibitory Input Pulse.
\newblock J Comput Neurosci. 2010;28(3):509--526.

\bibitem{AdaLeB17}
Adams NE, Sherfey JS, Kopell NJ, Whittington MA, LeBeau FEN.
\newblock Hetereogeneity in Neuronal Intrinsic Properties: A Possible Mechanism
  for Hub-Like Properties of the Rat Anterior Cingulate Cortex during Network
  Activity.
\newblock Eneuro. 2017;4(1):UNSP e0313--16.2017.
\newblock doi:{10.1523/ENEURO.0313-16.2017}.

\bibitem{WitBat13}
Witt A, Palmigiano A, Neef A, {El Hady} A, Wolf F, Battaglia D.
\newblock Controlling the oscillation phase through precisely timed closed-loop
  optogenetic stimulation: a computational study.
\newblock Front Neural Circuits. 2013;7:49.
\newblock doi:{10.3389/fncir.2013.00049}.

\bibitem{KirBat16}
Kirst C, Timme M, Battaglia D.
\newblock Dynamic information routing in complex networks.
\newblock Nature Communications. 2016;7:11061.
\newblock doi:{10.1038/ncomms11061}.

\bibitem{IacTsa16}
Iaccarino HF, Singer AC, Martorell AJ, Rudenko A, Gao F, Gillingham TZ, et~al.
\newblock Gamma frequency entrainment attenuates amyloid load and modifies
  microglia.
\newblock Nature. 2016;540(7632):230--+.
\newblock doi:{10.1038/nature20587}.

\bibitem{MarTsa19}
Martorell AJ, Paulson AL, Suk HJ, Abdurrob F, Drummond GT, Guan W, et~al.
\newblock Multi-sensory Gamma Stimulation Ameliorates Alzheimer's-Associated
  Pathology and Improves Cognition.
\newblock Cell. 2019;177:256--271.e22.
\newblock doi:{10.1016/j.cell.2019.02.014}.

\bibitem{MenRie18a}
Meng JH, Riecke H.
\newblock Synchronization by uncorrelated noise: interacting rhythms in
  interconnected neuronal networks.
\newblock BMC Neuroscience. 2018;19 (Suppl 2):116.

\bibitem{BrWa03}
Brunel N, Wang XJ.
\newblock What Determines the Frequency of Fast Network Oscillations With
  Irregular Neural Discharges? {I.} {Synaptic} Dynamics and
  Excitation-Inhibition Balance.
\newblock J Neurophysiol. 2003;90(1):415--430.

\end{thebibliography}

\newpage{}

\renewcommand{\thefigure}{S\arabic{figure}} \setcounter{figure}{0} 

\section*{Supplementary Information }

\begin{figure}[h]
\begin{centering}
\includegraphics[width=0.6\linewidth]{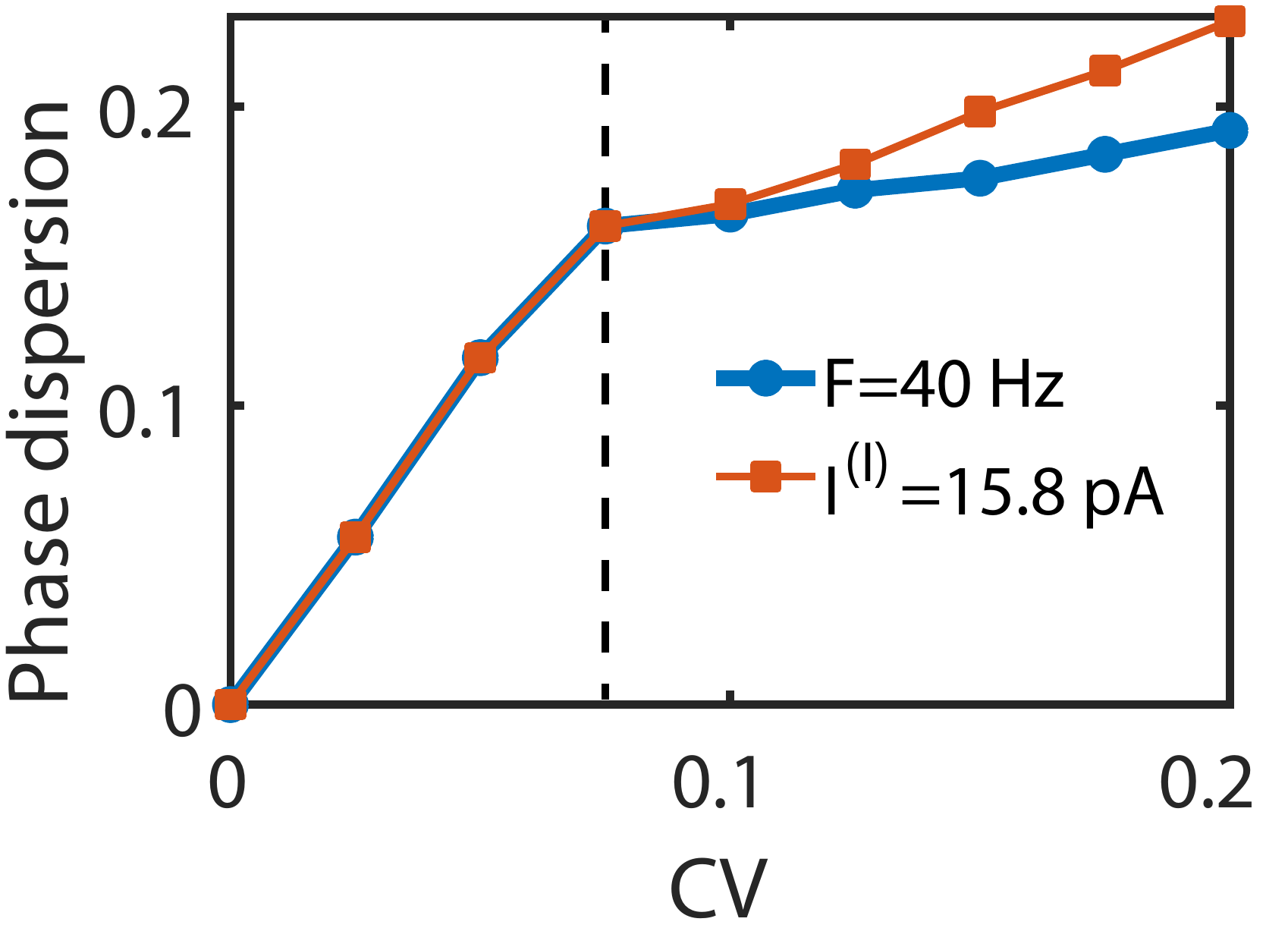}
\par\end{centering}
\caption{Dependence of the phase dispersion on the heterogeneity of the bias
current. The phase dispersion was determined as the time difference
between the first and the last spike in the same spike volley normalized
by the period. Blue: fixed natural frequency ($f_{network}=40$Hz)
for different neuronal heterogeneity. Red: fixed mean input strength
($I^{(I)}=$15.8 pA) for different neuronal heterogeneity. For $CV\protect\geq$
0.075 (dashed line), some neurons spike more than once in a cycle.
\label{fig:phase_dispersion_CV}}
\end{figure}

\newpage{}

\begin{figure}[H]
\begin{centering}
\includegraphics[width=0.6\linewidth]{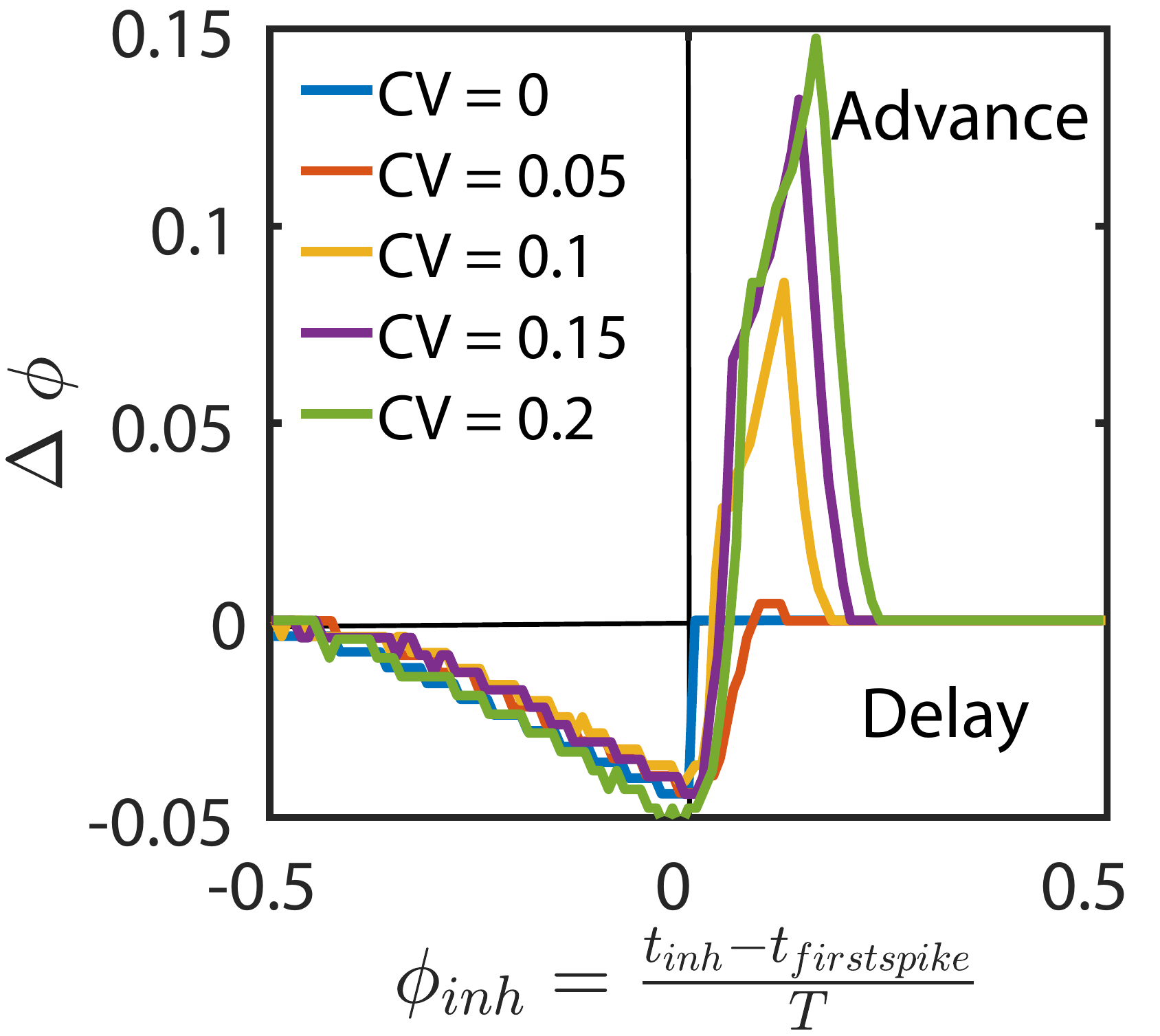}
\par\end{centering}
\caption{fmPRC of heterogeneous ING networks for fixed steady current ($I^{(I)}=$15.8
pA) instead of fixed frequency (cf. Fig.\ref{fig:Heterogeneous-network-biphasic}A).
The paradoxical phase advance increased with neuronal heterogeneity.
\label{fig:fixed_tonic_current}}
\end{figure}

\newpage{}
\begin{figure}[H]
\begin{centering}
\includegraphics[width=1\linewidth]{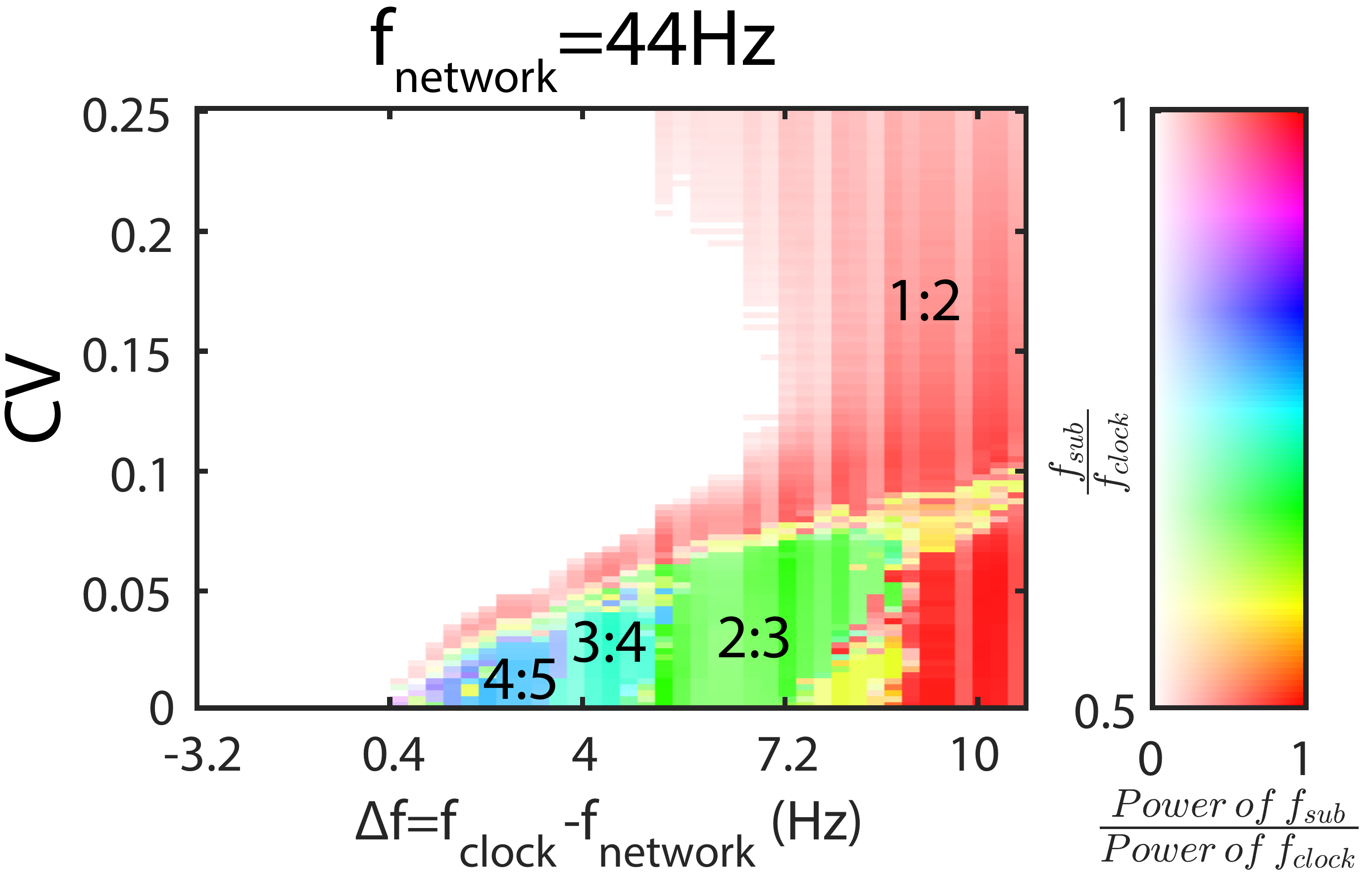}
\par\end{centering}
\caption{Subharmonic response of the ING rhythm with a longer synaptic delay
within the network ($\tau_{d}=5$ ms) receiving periodic inhibitory
input. For each value of the input heterogeneity, the natural frequency
$f_{network}$ was kept constant ($f_{network}=44$ Hz) by adjusting
the mean input\textcolor{black}{{} strength $I^{(I)}$.} The range of
detuning where increasing heterogeneity induced a type 3 synchronization
became wider compared to Fig.\ref{fig:fmPRC-syn}B, where $\tau_{d}=3$
ms. $W^{(ext)}=1.2\times10^{-3}$. \label{fig:fmPRC-syn-longerdelay}}
\end{figure}

\end{document}